\documentclass[12pt]{article}
\usepackage{latexsym,amsmath,amssymb}
\usepackage{graphicx}

\begin{document}

%\begin{flushright}
%Sofia University\\
%\end{flushright}
%%%%%%%%%%%%%%%%%%%%%%%%%%%%%%%%%%%%%%%%%%%%%%%%%%%%%%%%%%%%%%%%%%%

\title{Charged anti-de Sitter scalar-tensor black holes and their thermodynamic phase structure }

\author{Daniela D. Doneva$^{1,2}$\thanks{E-mail: ddoneva@phys.uni-sofia.bg}\,,\:  Stoytcho S. Yazadjiev$^{3}$ \thanks{E-mail:
yazad@phys.uni-sofia.bg}
\,,\: Kostas D. Kokkotas$^{2,4}$ \thanks{E-mail: kostas.kokkotas@uni-tuebingen.de} \,\,\\{\footnotesize  ${}^{1}$Dept.
of Astronomy,
                Faculty of Physics, St.Kliment Ohridski University of Sofia}\\
                {\footnotesize  5, James Bourchier Blvd., 1164 Sofia, Bulgaria }\\\\[-3.mm]
      {  \footnotesize ${}^{2}$ Theoretical Astrophysics, Eberhard-Karls University of T\"ubingen, T\"ubingen 72076, Germany }\\
       {  \footnotesize ${}^{3}$Dept. of Theoretical Physics,
                Faculty of Physics, St.Kliment Ohridski University of Sofia}\\
{\footnotesize  5, James Bourchier Blvd., 1164 Sofia, Bulgaria }\\
  {\footnotesize ${}^{4}$ Department of Physics, Aristotle University of Thessaloniki, Thessaloniki 54124, Greece}\\
\\[-3.mm]
 Ivan Zh. Stefanov$^{5}$ \thanks{E-mail: izhivkov@yahoo.com}\,,\:   Michail D.~Todorov$^{6}$\thanks{E-mail: mtod@tu-sofia.bg}
\\ [2.mm]{\footnotesize{${}^{5}$ Department of Applied Physics, Technical University of Sofia,}}\\ [-1.mm]{\footnotesize{8, Kliment Ohridski Blvd., 1000
Sofia, Bulgaria}}
\\\\[-3.mm]{\footnotesize
{${}^{6}$Faculty of Applied Mathematics and Computer Science, Technical University of Sofia}}\\
[-1.mm] {\footnotesize 8, Kliment Ohridski Blvd., 1000 Sofia, Bulgaria}}

\date{}

\maketitle

\begin{abstract}
In the present paper we numerically construct new charged anti-de Sitter black holes coupled to
nonlinear Born-Infeld electrodynamics within a certain class of scalar-tensor theories. The properties of the solutions are investigated both numerically and analytically.
We also study the thermodynamics of the black holes in the canonical ensemble. For large values of the Born-Infeld parameter and
for a certain interval of the  charge values we find the existence of a first-order phase transition between small and very large black holes.
An unexpected result is that for a certain small charge subinterval two phase transitions have been observed, one of zeroth and one of first order.
It is important to note that such phase transitions are also observed for pure Einstein-Born-Infeld-AdS black holes.

\end{abstract}

%%%%%%%%%%%%%%%%%%%%%%%%%%%%%%%%%%%%%%%%%%%%%%%%%%%%%%%%%%%%%%%%%%%

%\draft
\sloppy

\section{Introduction}
In the last decade  anti-de Sitter (AdS) black holes and especially  their thermodynamics have attracted considerable interest
due to the  AdS/CFT duality. According to the   AdS/CFT  conjecture the thermodynamics of the AdS black holes is related to the thermodynamics
of the dual conformal field theory (CFT) residing on the boundary of the AdS space \cite{Maldacena, Petersen}. In  their pioneering work Hawking and Page \cite{HP} showed the existence of a phase
transition between the AdS black hole and thermal AdS space. The AdS/CFT duality provides us with a new tool to study
the phase transitions in the dual CFT theories on the base of studying the phase transitions of the AdS black holes.

Charged AdS black holes and their thermodynamics have been studied not only within the linear Maxwell electrodynamics but also within the framework of
the nonlinear electrodynamics of Born-Infeld.  This nonlinear electrodynamics was first introduced by Born and Infeld in 1934 as an attempt to obtain
a finite energy density model for the electron \cite{BI}. The interest for nonlinear electrodynamics has been later revived in the context of string theory. It arises naturally in open strings and D-branes \cite{nle_str1} -- \cite{L}. Nonlinear electrodynamics
models coupled to gravity have been discussed in different aspects (see, for example, \cite{Wil} -- \cite{Breton2} and references therein). In
particular,  black holes coupled to Born-Infeld electrodynamics, including in the presence of a cosmological constant, have been studied in numerous
papers \cite{TT2},\cite{Garcia} -- \cite{Sheykhi_topol}, and their thermodynamics in \cite{Fern} -- \cite{Sheykhi_TD_tpol}.

The aim of the present paper is to study the charged AdS black holes coupled to nonlinear Born-Infeld electrodynamics and their thermodynamics within
the framework of a certain class of  scalar-tensor theories (STT). Scalar-tensor theories of gravity are among the most natural
generalizations of general relativity (GR) and arise  in string theory and higher dimensional gravity theories
\cite{Will}. The fundamental question is whether objects found and studied in the frame of GR, such as black holes, would have different properties
in the frame of the scalar-tensor theories.
Recent studies on asymptotically flat, scalar-tensor black holes coupled to nonlinear electrodynamics show that the presence of a scalar field in the gravitational sector
leads to interesting and serious consequences for the black holes. In the general case the scalar field restricts the possible
causal structures in comparison to the pure Einstein gravity \cite{SYT1, SYT2}.  For some classes of scalar-tensor theories \cite{SYT3} there even
exist
non-unique scalar-tensor black hole solutions with the same conserved asymptotic charges. In other words the spectrum of
asymptotically flat black hole solutions coupled to nonlinear electrodynamics in the STT is much  richer and much more complicated than in general
relativity. This shows that the scalar-tensor black holes coupled to nonlinear electrodynamics and other sources with non-vanishing trace of
the energy-momentum tensor (in four dimensions) constitute an area that deserves further study.
In view of the  current great interest in the AdS black holes, the natural
next step is to study the scalar-tensor black holes coupled to nonlinear electrodynamics in spacetimes with AdS asymptotic.

In  this paper we construct new numerical solutions describing charged AdS scalar-tensor black holes coupled to nonlinear
Born-Infeld electrodynamics and we make a parameter study of their properties. The thermodynamics of the constructed solutions in canonical ensemble is
also studied. We found that for large values of the Born-Infeld parameter $b$
and for a certain interval of intermediate values of the charge there exist a first-order phase transition between small and very large black holes.
An interesting result is that for a small subinterval
one more phase transition exists which is of zeroth order since the thermodynamic potential is
discontinuous there -- it jumps to a lower value. Such phase transition of first and zeroth order can be observed also in pure Einstein gravity,
i.e. when no scalar field is present.

The phase structure of the Einstein-Born-Infeld black
holes and the possible Hawking-Page transitions have been examined in \cite{Fern} and \cite{Myung3}. Hawking-Page phase transitions
have been studied for different black holes, for example, Einstein-Maxwell black holes in spacetime with different dimensions \cite{EmparanJohnson, EmparanJohnson1},
in three-dimensional spacetime \cite{Myung1, Myung2}, in higher-derivative gravity \cite{Dey1, Dey2}, in higher-curvature gravity
\cite{Cho}, and
in unusual electrodynamics which does not restore the Maxwell electrodynamics in the weak-field limit \cite{MGH}.
Phase transitions between solitons and black holes in asymptotically AdS/$\mathbb{Z}_k$ spaces have been studied in \cite{MannStotyn}.

\section{Formulation of the problem}
The most general form of the action in scalar-tensor theories is
\begin{align}
S=\frac{1}{16\pi G_*} \int{d^4 x \sqrt{-\tilde{g}}\left(F(\Phi)\tilde{R} - Z(\Phi)\tilde{g}^{\mu\nu}\partial_\mu \Phi \partial_\nu \Phi -
2U(\Phi)\right)} + S_m[\Psi_m;\tilde{g}_{\mu\nu}], \label{Action_JF}
\end{align}
where $G_*$ is the bare gravitational constant and $\tilde{R}$ is the Ricci scalar curvature with respect to the spacetime metric $\tilde{g}^{\mu\nu}$.
$F(\Phi)$, $Z(\Phi)$, and $U(\Phi)$ are functions of the scalar field $\Phi$ and their specific choice determines the scalar-tensor theory completely.
In order for the gravitons to carry positive energy the function $F(\Phi)$ must be positive ($F(\Phi)>0$), while the non-negativity of the scalar field energy requires
that $2F(\Phi)Z(\Phi) + 3[dF(\Phi)/d\Phi]^2\geq 0$. The action of the  sources is $S_m[\Psi_m;\tilde{g}_{\mu\nu}]$ and there is no direct coupling
between the sources of gravity and the scalar field in order for the weak equivalence principal to be satisfied.

The action (\ref{Action_JF}) is in the so called Jordan frame which is the physical frame but it is more convenient to work in the Einstein frame. The
relations between the two frames are
\begin{align}\label{CONFTRANS}
&g_{\mu\nu}=F(\Phi) \tilde{g}_{\mu\nu}, \\ \notag \\
&\left(\frac{d\varphi}{d\Phi}\right)^2=\frac{3}{4}\left(\frac{d \ln(F(\Phi))}{d\Phi}\right)^2 + \frac{Z(\Phi)}{2F(\Phi)},
\end{align}
where $g_{\mu\nu}$ and $\varphi$ are correspondingly the metric and the scalar field in the Einstein frame. The additional
introduction of the functions
\begin{align}
&{\cal A} (\varphi) = F^{-1/2}(\Phi), \\
&V(\varphi) = \frac{1}{2} U(\Phi) F^{-2} (\Phi),
\end{align}
leads to the following action in the Einstein frame:
\begin{align}
S=\frac{1}{16\pi G_*} \int{d^4 x \sqrt{-g}\left(R - 2g^{\mu\nu}\partial_\mu \varphi \partial_\nu \varphi - 4V(\varphi)\right)} + S_m[\Psi_m;{\cal
A}^2(\varphi) g_{\mu\nu}], \label{Action_EF}
\end{align}
where $R$ is the Ricci scalar curvature with respect to the spacetime metric $g_{\mu\nu}$. In the Einstein frame there is direct coupling between the
sources of gravity and the scalar fields through the coupling function ${\cal A}(\varphi)$ and the specific choice of the scalar-tensor theory is
completely determined by this function and by the potential of the scalar field $V(\varphi)$.

Here we will consider nonlinear electrodynamics and its action in the Einstein frame is given by
\begin{align}
S_m = \frac{1}{4\pi G_*} \int{d^4 x \sqrt{-g} {\cal A}^4(\varphi)L(X,Y) }, \label{Actiona_BI}
\end{align}
where $L(X,Y)$ is the Lagrangian of the nonlinear electrodynamics. The equations defining  the functions $X$ and $Y$ are
\begin{align}
    &X=\frac{{\cal A}^{-4}(\varphi)}{4}F_{\mu\nu}F^{\mu\nu}, \label{f:eqX}\\ \notag \\
    &Y=\frac{{\cal A}^{-4}(\varphi)}{4}F_{\mu\nu}(\star F)^{\mu\nu},
\end{align}
where $F_{\mu\nu}$ is the electromagnetic field strength tensor and $\star$ stands for the Hodge dual with respect to the metric $g_{\mu\nu}$.

The Lagrangian of the Born-Infeld nonlinear electrodynamics is
\begin{align}
L = 2b\left[1-\sqrt{1+\frac{X}{b} - \frac{Y^2}{4b^2}}\right],
\end{align}
where $b$ is a parameter and in the limit $b\rightarrow \infty$ the linear  Maxwell electrodynamics is restored.

The variation of the action (\ref{Action_EF}) leads to the following field equations
\begin{eqnarray}
&&R_{\mu\nu} = 2\partial_{\mu}\varphi \partial_{\nu}\varphi +  2V(\varphi)g_{\mu\nu} -
 2\partial_{X} L(X, Y) \left(F_{\mu\beta}F_{\nu}^{\beta} -
{1\over 2}g_{\mu\nu}F_{\alpha\beta}F^{\alpha\beta} \right)  \nonumber \\
&&\hspace{2cm}-2{\cal A}^{4}(\varphi)\left[L(X,Y) -  Y\partial_{Y}L(X, Y) \right] g_{\mu\nu}, \nonumber  \\ \nonumber \\
&&\nabla_{\mu} \left[\partial_{X}L(X, Y) F^{\mu\nu} + \partial_{Y}L(X, Y) (\star F)^{\mu\nu} \right] = 0 \label{F},\\ \nonumber \\
&&\nabla_{\mu}\nabla^{\mu} \varphi = {d V(\varphi)\over d\varphi } -
4\alpha(\varphi){\cal A}^{4}(\varphi) \left[L(X,Y) -  X\partial_{X}L(X,Y) -  Y\partial_{Y}L(X, Y) \right], \nonumber
\end{eqnarray}
where
\begin{equation}
\alpha(\varphi) = {d \ln{\cal A}(\varphi)\over d\varphi}. \label{alpha_def}
\end{equation}

In the present paper we  are interested in  spacetimes with anti-de Sitter asymptotic structure and thus we will consider the following type of
potential:
\begin{align}\label{EFPOTENTAIL}
V=\frac{1}{2}\Lambda,
\end{align}
where $\Lambda<0$ is a constant. The coefficient $1/2$ is a standard normalizing factor.

Let us explain the reasons for choosing the potential (\ref{EFPOTENTAIL}). For this purpose we consider the Jordan frame and we require the metric
to have AdS asymptotic\footnote{We mean AdS asymptotic corresponding to a cosmological term $\Lambda$.} in this frame. Then we have two possibilities -- either the factor $F(\Phi)$ is finite at infinity
(i.e. $0<F(\Phi_{\infty})<\infty$)   or it is divergent at infinity.
Although the case when $F(\Phi)$ is divergent  at infinity  could be of some interest, it is not generic and in fact is  degenerate since it corresponds  to zero effective gravitational constant at infinity. Here we consider only "physically well-behaved" solutions
with finite\footnote{Throughout this paper, without loss of generality, we set $F(\Phi_{\infty})=1$.} $F(\Phi_{\infty})>0$ . Using the (Jordan frame) field equations and especially the equation for $\Phi$, one can show that the finiteness of $F(\Phi)$ at infinity  requires\

\begin{eqnarray}\label{PotentialCon}
&&\lim_{\Phi\to \Phi_{\infty}} \left[\frac{\frac{dU(\Phi)}{d\Phi}F(\Phi) - 2U(\Phi)\frac{dF(\Phi)}{d\Phi}}{2Z(\Phi)F(\Phi) + 3\left(\frac{dF(\Phi)}{d\Phi}\right)^2}\right]
=\lim_{\Phi\to \Phi_{\infty}}
\left[\frac{F^{\,3}(\Phi) \frac{d}{d\Phi}\left(\frac{U(\Phi)}{F^{2}(\Phi)}\right)}{2Z(\Phi)F(\Phi) + 3\left(\frac{dF(\Phi)}{d\Phi}\right)^2}
 \right]= 0,  \nonumber\\ \\
&&\lim_{\Phi\to \Phi_{\infty}} U(\Phi)=\Lambda . \nonumber
\end{eqnarray}
The simplest (Jordan frame) potential satisfying this condition\footnote{More precisely, we mean the simplest potential in the framework
of the class of scalar-tensor theories considered in the present paper. For this class of scalar-tensor theories the
denominator in (\ref{PotentialCon}) is finite. }, i.e. allowing simultaneously  AdS asymptotic for the metric and
finite $F(\Phi_{\infty})>0$, is $U(\Phi)=\Lambda F^{2}(\Phi)/F^{2}(\Phi_{\infty})$  which corresponds exactly to the Einstein frame potential
(\ref{EFPOTENTAIL}) for $F(\Phi_{\infty})=1$. The conditions
(\ref{PotentialCon}) written  in  the Einstein frame take the form

\begin{eqnarray}
\lim_{\varphi\to\varphi_{\infty}}\frac{dV(\varphi)}{d\varphi}=0, \;\;\;\;\; \lim_{\varphi\to\varphi_{\infty}} V(\varphi)=\frac{1}{2}\Lambda.
\end{eqnarray}
From here it is obvious  that $V(\varphi)=\Lambda/2$ is indeed the simplest choice of a potential admitting the desired properties.

One can show, as we shall see below, that for the scalar-tensor theories under consideration,  the asymptotic behavior of the factor $F(\Phi)$ is $F(\Phi)\approx 1 +  const/r^3$  and it guarantees the same anti-de Sitter asymptotic structure simultaneously in both the Jordan and the Einstein  frames. Moreover, the asymptotic
behavior of $F(\Phi)$ guarantees that the mass of the Jordan frame solution is the same as that of the Einstein frame solution. This can be easily
seen from the results in Appendix \ref{TD_action}.

Another reason for considering finite $F(\Phi_{\infty})$ comes from the thermodynamics. The black hole thermodynamics in scalar-tensor theories is
naturally defined in the Einstein frame as  discussed in Appendix \ref{STTERMODYNAMICS}. When $F(\Phi)$ is divergent at infinity the conformal transformation
does not preserve the AdS asymptotic  when we move from the Jordan to the Einstein frame. In this case  the expected thermodynamic would be rather
different from that of the AdS black holes if it could be defined at all because of the rather unusual spacetime asymptotic.

\section{Basic equations}
\subsection{The reduced system}
Here and below we will work in a system of units in which $G_*=c=\mu_0/4\pi=1$, where $\mu_0$ is the magnetic permeability of the vacuum. In this system
$$
[P]=m; \,\,\,\,[M]=m;\,\,\,\,[b]=m^{-2}.
$$
We will consider a static and spherically symmetric spacetime and the ansatz for the metric (in the Einstein frame) is
\begin{equation}
ds^2 = g_{\mu\nu}dx^{\mu}dx^{\nu} = - f(r)e^{-2\delta(r)}dt^2 + {dr^2\over f(r) } +
r^2\left(d\theta^2 + \sin^2\theta d\phi^2 \right). \label{metric}
\end{equation}

An important property of the Born-Infeld electrodynamics is the electric-magnetic duality symmetry of the theory \cite{GR1} -- \cite{Ferrara}.
It is sufficient to study
only the magnetically charged case, and the electrically charged solution can be obtained from the magnetically charged solution using the electric-magnetic
rotation defined by
\begin{equation}
\{ g_{\mu\nu},\, \varphi,\, F_{\mu\nu},\, P ,\,X,\, L(X)\}  \longleftrightarrow \
\{ g_{\mu\nu},\, \varphi,\, ~\star
G_{\mu\nu},\, \bar{Q} ,\,\bar{X},\, L(\bar{X})\},\label{duality}
\end{equation}
where the functions denoted by $\bar{(..)}$ correspond to the dual solution, $P$ is the magnetic charge, $\bar{Q}$ is the charge of the dual
solution,
and
\begin{equation}
G_{\mu\nu}=-2\frac{\partial \left[{\cal A}^{4}(\varphi)L\right]}{\partial F^{\mu\nu} }, \mbox{ } \bar{X}=-\Bigl[\partial_{X}L(X)\Bigr]^2 X.
\end{equation}

In the magnetically charged case the electromagnetic field strength tensor is
\begin{equation}
F = P \sin\theta d\theta \wedge d\phi,
\end{equation}
where $P$ is the magnetic charge. For the functions $X$ and $Y$ we obtain
\begin{equation}
X = {{\cal A}^{-4}(\varphi)\over 2} \frac{P^2}{r^4}, \mbox{ } Y=0.
\end{equation}
The truncated Born-Infeld Lagrangian is
\begin{equation}
L(X) = 2b \left( 1- \sqrt{1+ \frac{X}{b}} \right)\label{LBI}.
\end{equation}

Using the metric (\ref{metric}) the field equations reduce to the following system of coupled ordinary differential equations

\begin{eqnarray}
&&f'' -2f\delta'' -3f'\delta'+
2f\delta'^2+\frac{2}{r}f'-\frac{4}{r}f\delta' =\nonumber\\
&&\,\,\,\,\,\,\,\,\,\,\,\,\,\,\,\,\,\,\,\,\,\,\,\,\,\,\,\,\,\,\,\,\,\,\,\,\,\,\,\,\,\,\,\,\,\,\,\,\,\,-4\left\{V(\varphi)+{\cal A}^{4}(\varphi)
\left[2X\partial_{X}L(X)-L(X)\right]\right\}\label{EQ_tt},\\\nonumber\\
&&f'' -2f\delta'' -3f'\delta'+2f\delta'^2+\frac{2}{r}f'=\nonumber\\
&&\,\,\,\,\,\,\,\,\,\,\,\,\,\,\,\,\,\,\,\,\,\,\,\,\,\,\,\,\,\,\,\,\,\,\,\,\,\,\,\,\,\,\,\,\,\,\,\,\,\,-4\left\{f\varphi\,'^{\,2}+V(\varphi)+{\cal
A}^{4}(\varphi) \left[2X\partial_{X}L(X)-L(X)\right]\right\}\label{EQ_rr},\\ \nonumber\\
&&1-f-rf'+rf\delta'=2r^2\left[ V(\varphi)-
{\cal A}^{4} (\varphi)L(X)  \right] \label{EQ_thita},\\ \nonumber\\
&&\frac{d }{dr}\left( e^{-\delta}r^{2}f\frac{d\varphi }{dr} \right)=\frac{d V(\varphi)}{d\varphi}e^{-\delta}r^2+4 r^2 e^{-\delta} \alpha(\varphi)
{\cal A}^{4}(\varphi)\left[X\partial_{X}L(X)- L(X)\right]  \label{EQPhi_d}   .
\end{eqnarray}
These are four equations for only three unknown functions $f$, $\varphi$ and $\delta$ but the self-consistency of the system is guaranteed by the
Bianchi identity.

In pure Einstein theory the solution describing Schwarzschild-AdS black holes is
\begin{eqnarray}
&&\delta_{\rm E}(r)=0, \\
&&f_{\rm E}(r)=1-\frac{2M_{\rm E}}{r} - \frac{1}{3} \Lambda r^2,
\end{eqnarray}
where $M_{\rm E}$ is the mass of the black hole. The asymptotic of the metric function $f(r)$ at infinity in our problem is the same
$(- \frac{1}{3} \Lambda r^2)$ and this means that
$f(r)$ is unbounded at infinity. From a numerical point of view it is convenient to introduce a new unknown function $m(r)$ (which is finite at infinity
in the considered class of theories) using the substitution
\begin{eqnarray}\label{localmass}
f(r)=1-\frac{2m(r)}{r} - \frac{1}{3} \Lambda r^2 \label{f_eq}.
\end{eqnarray}

Using the system (\ref{EQ_tt})-(\ref{EQPhi_d}) after some manipulations
we can obtain the following simpler system of ordinary differential equations
\begin{eqnarray}
&&\frac{d\delta}{dr}=-r\left(\frac{d\varphi}{dr} \right)^2 \label{EQDelta},\\
&&\frac{d m}{dr}=r^2\left[\frac{1}{2}f\left(\frac{d\varphi}{dr} \right)^2 -
{\cal A (\varphi)}^{4}L(X)  \right] \label{EQm},\\
&&\frac{d }{dr}\left( r^{2}f\frac{d\varphi }{dr} \right)=
r^{2}\left\{-4\alpha(\varphi){\cal A}^{4}(\varphi) \left[L -  X\partial_{X}L(X)\right] -
r f\left(\frac{d\varphi}{dr} \right)^3    \right\} \label{EQPhi}  .
\end{eqnarray}
The first equation (\ref{EQDelta}) is decoupled and it can be solved separately once a solution for the functions $m(r)$ and $\varphi(r)$ is obtained.

We will consider a class of scalar-tensor theories for which $\alpha(\varphi)=\mathrm{const}\equiv\alpha>0$\footnote{Constant $\alpha(\varphi)$
corresponds to Brans-Dicke scalar-tensor theory.}.
The investigation of the case with  $\alpha(\varphi)=\mathrm{const}\equiv\alpha<0$ is similar. Although we restrict ourselves to
$\alpha(\varphi)=\mathrm{const}\equiv\alpha$ the obtained results are
qualitatively\footnote{As the numerical results show the picture is not only  qualitatively but also quantitatively close for
 all scalar-tensor theories with the same $\alpha(0)$.} the same even for more general coupling functions $\alpha(\varphi)>0$
 for which $\alpha(0)=\alpha$.

In the chosen ansatz for the metric (\ref{metric}) the temperature is given by the following relation
\begin{equation}
T=\left.{ f'(r)\,e^{-\delta(r)}\over 4\pi}\right|_{r=r_H}.
\end{equation}

\subsection{Asymptotic behavior}
 The asymptotic behavior of the functions $\delta$, $m$ and $\varphi$ at $r\rightarrow\infty$ is
\begin{eqnarray}
&&\delta|_{r \rightarrow \infty}=\frac{3C_3^2}{2r^6} + {\cal O}(r^{-8}), \label{delta_asymptotic} \\
&&m|_{r \rightarrow \infty} = M - \frac{P^2}{2 r} + \frac{\Lambda C^2_3}{2 r^3} + {\cal O}(r^{-5}), \label{m_asymptotic} \\ &&\notag\\
&&\varphi|_{r \rightarrow \infty} = \frac{C_3}{r^3} + \frac{9C_3}{5\Lambda r^5} - \frac{3 M C_3}{\Lambda r^6} + {\cal
O}(r^{-7}),  \label{phi_asymptotic}
\end{eqnarray}
where $C_3$ is a constant and $M$ is the mass of the black hole in the Einstein frame (for the definition of the mass in the AdS spaces see
Appendix A). Let us compare
the asymptotic behavior of these functions in AdS and in asymptotically flat spacetime. In AdS spacetime the scalar field decreases as $1/r^3$ and
this is much faster than in the asymptotically flat case where the scalar field decreases as $1/r$. The function $\delta(r)$ also decreases much
faster in AdS spacetime, where the leading term is proportional to $1/r^6$ compared to the leading term $1/r^2$ in the asymptotically flat spacetime. The
asymptotic behavior of  $m(r)$ up to first order in $1/r$ is the same.

\subsection{Qualitative properties}\label{3.3}
We can obtain some information for the black hole solutions using equations (\ref{EQPhi_d}) and (\ref{EQDelta}) and the boundary conditions.
We prove that the black holes under consideration have simpler causal structure than the black holes in pure Einstein-Born-Infeld theory in
AdS spacetime and we also obtained some information for the behavior of the unknown functions. The general
properties of the considered scalar-tensor black holes coupled to Born-Infeld nonlinear electrodynamics in AdS spacetime can be summarized as follows:
\begin{itemize}
\item[1.] The black holes have only one event horizon and extremal black holes do not exist.

\item[2.] The scalar field $\varphi$ is negative on the event horizon, it has no zeros and increases  monotonically.

\item[3.] The metric function $\delta$  is positive on the event horizon, it has no zeros and decreases  monotonically.
\end{itemize}
To prove these statements we should first note that it can be easily shown that for the Born-Infeld
electrodynamics the following inequality is satisfied:
\begin{equation}
 X\partial_{X}L(X)- L(X)>0\label{EDHAM}.
\end{equation}

We will first prove that the considered black holes have only one horizon. Let us assume that they have two  horizons $r_-$ and $r_+$,
where $r_+ > r_-$.  When we integrate equation (\ref{EQPhi_d}) in the interval $r\in[r_{-},r_{+}]$ we obtain

\begin{eqnarray}
&&\left. \left( e^{-\delta}r^{2}f\frac{d\varphi }{dr} \right) \right|_{r_{+}}-
\left.\left( e^{-\delta}r^{2}f\frac{d\varphi }{dr} \right) \right|_{r_{-}}=\nonumber \\
&&=4 \int\limits_{r_{-}}^{r_{+}}r^2 e^{-\delta} \alpha(\varphi)
{\cal A}^{4}(\varphi)\left[X\partial_{X}L(X)- L(X)\right] dr>0.
\end{eqnarray}
The horizons are defined as the points where $f(r)=0$ so the left-hand side of the equation is zero. But from (\ref{EDHAM}) and the fact
that we consider $\alpha(\varphi)>0$ it is obvious that the right-hand side is strictly positive. So we reached a contradiction and this means that two black hole horizons cannot exist.

In order to prove that an extremal black hole cannot exist we will examine again equation (\ref{EQPhi_d}), that is
\begin{eqnarray}
&&\frac{d }{dr}\left( e^{-\delta}r^{2}f\frac{d\varphi }{dr} \right)=4 r^2 e^{-\delta} \alpha(\varphi)
{\cal A}^{4}(\varphi)\left[X\partial_{X}L(X)- L(X)\right]>0  \notag  .
\end{eqnarray}
Let us assume that we have a degenerate event horizon at $r_d$. The right-hand side of the equation is strictly positive for every
$r_d>0$,
\footnote{We are not interested in the case $r_d=0$ because this corresponds to naked singularity.} but the left-hand side is zero and we reach a contradiction.

We can investigate the behavior of the function $\varphi(r)$ in the domain $r\in[r_H,\infty)$ ($r_H$ is the radius of the event horizon) using
equation (\ref{EQPhi_d}). As we already noted the right-hand side of (\ref{EQPhi_d}) is positive. Integration of equation  (\ref{EQPhi_d})
from $r_H$ to $r>r_H$ gives
\begin{eqnarray}\label{Phi_realation}
e^{-\delta}r^2 f(r) \frac{d\varphi }{dr} = 4\int^{r}_{r_H}r^2e^{-\delta}\alpha(\varphi){\cal A}^{4}(\varphi)
\left[X\partial_{X}L(X) - L(X)\right]dr >0.
\end{eqnarray}
This shows that $d\varphi/dr>0$ in the domain $(r_H, \infty)$.  Therefore $\varphi (r)$ increases monotonically. Since the
asymptotic value of the scalar field at infinity is zero we can conclude that $\varphi$ should be negative on the horizon and monotonically
increasing to zero.
From (\ref{Phi_realation}) taking the limit $r\to \infty$ we find that $C_3<0$.

Finally, $\delta(r)$ decreases monotonically since the right-hand side of equation (\ref{EQDelta}) is always negative. The asymptotic value
of $\delta(r)$ at infinity is zero so we can conclude that it is positive on the event horizon and monotonically decreasing to zero.

All of the  above presented propositions remain valid for different nonlinear electrodynamics for which the relation (\ref{EDHAM}) holds.

\subsection{Dimensionless quantities}

The reduced field equations (\ref{EQDelta})--(\ref{EQPhi}) are invariant under the rigid rescaling

\begin{equation}
r \to \lambda r, \,\;\,  m \to \lambda m , \,\,\; \Lambda \to \lambda^{-2} \Lambda, \, \,\; b\to \lambda^{-2} b, \, \; P \to \lambda P.
\end{equation}
Therefore, one may generate in this way a family of new solutions. Thus, the mass, the temperature and the entropy of the  new solutions are given by
the formulas

\begin{equation}
M \to \lambda M , \,\, \; T \to \lambda^{-1} T, \,\, \; S \to \lambda^{2} S.
\end{equation}

As a consequence of the rigid symmetry we may restrict our study to the case $\Lambda=-1$ by introducing the dimensionless
quantities

\begin{eqnarray}
&&r \to \frac{r}{l}, \, \,\, m \to \frac{m}{l}, \,\,\, b\to bl^2 , \,\,\, P\to \frac{P}{l}, \nonumber \\
&&M \to \frac{M}{l}, \,\,\, T\to l T, \,\,\, S\to \frac{S}{l^2},
\end{eqnarray}
where the choice of $\Lambda=-1$ requires

\begin{eqnarray}
l=\frac{1}{\sqrt{-\Lambda}}.
\end{eqnarray}

\subsection{Posing the Boundary-Value Problem}
The domain of integration is half-infinite $r\in[r_H,\infty)$ where the horizon, which is the left boundary,
is \emph{a priori} unknown.
We impose the following boundary conditions:

--- At infinity:
\begin{eqnarray}
&&\lim_{r \to \infty}m(r) =M, \label{cond_delta} \\
&&\lim_{r \to \infty}\delta(r)=\lim_{r \to \infty}\varphi(r)=0\label{cond_phi_delta},
\end{eqnarray}
where $M$ is the mass of the black hole in the Einstein frame (see Appendix A);

--- On the horizon:
\begin{equation}
f(r_H)=0.\label{cond_horizon}
\end{equation}

The following regularization condition should also be fulfilled on the horizon:
\begin{equation}
\left.\frac{df}{dr}\!\cdot\! \frac{d \varphi}{d r}-\left\{ 4 \alpha(\varphi) {\cal A}^4(\varphi) [X \partial_X L(X)-L(X) ]\right\}\right|_{r=r_H}=0.
\label{regul}
\end{equation}

In this way the governing equations \eqref{EQDelta}--\eqref{EQPhi} together with the above boundary conditions  \eqref{cond_delta}--\eqref{regul}, complement
a well-posed boundary value problem. Since the position of the event horizon (the left boundary of the integration interval) is not fixed, moreover it is
unknown\footnote{Such BVPs are known in mathematical physics as BVPs of the Stefan kind.} and there is no analytic way to determine it,
we can use one of these boundary conditions to determine it.
We can introduce a new shifted independent variable  $x=r-r_H$. In this way the domain of integration is completely defined -- $x \in [0,\infty)$
and the radius of the horizon will appear explicitly in the equations as an unknown parameter. So, we have one parametric problem, which can be
considered as a spectral-like nonlinear BVP.

Let us emphasize that there is an alternative formulation of the problem where the horizon is an input parameter. In this case we do not need a
shifting map and the mass $M$  becomes the sought parameter. We used this method of solution for justification of our results.

\subsection{Method of solution}
The system of ordinary differential equations (\ref{EQDelta})--(\ref{EQPhi}) is coupled and
nonlinear. It does not admit a global analytic solution and only some local expansions are possible. This is the reason to solve it numerically
by using the reliable Continuous Analog of Newton's Method, which is distinguished by its quadratic convergence in the vicinity of a localized
root \cite{gavurin, jidkov}\footnote{Let us note that in our case the roots appear to be functions in Banach space.}. Yet, the method has been
already successfully applied to similar problems (see, for example, \cite{JGGA2,  SYT1, SYT2, SYT3}).

\section{Discussion of numerical results}\label{sect_num}
As we noted in the previous sections we will consider constant parameter $\alpha>0$ ($\alpha$ is defined by equation (\ref{alpha_def})). The
results do not change qualitatively for different $\alpha>0$ and we will present the results for $\alpha=0.01$.
\footnote{This value is consistent with the current observational data \cite{Will}.}

\begin{figure}[htbp]%
\includegraphics[width=0.50\textwidth]{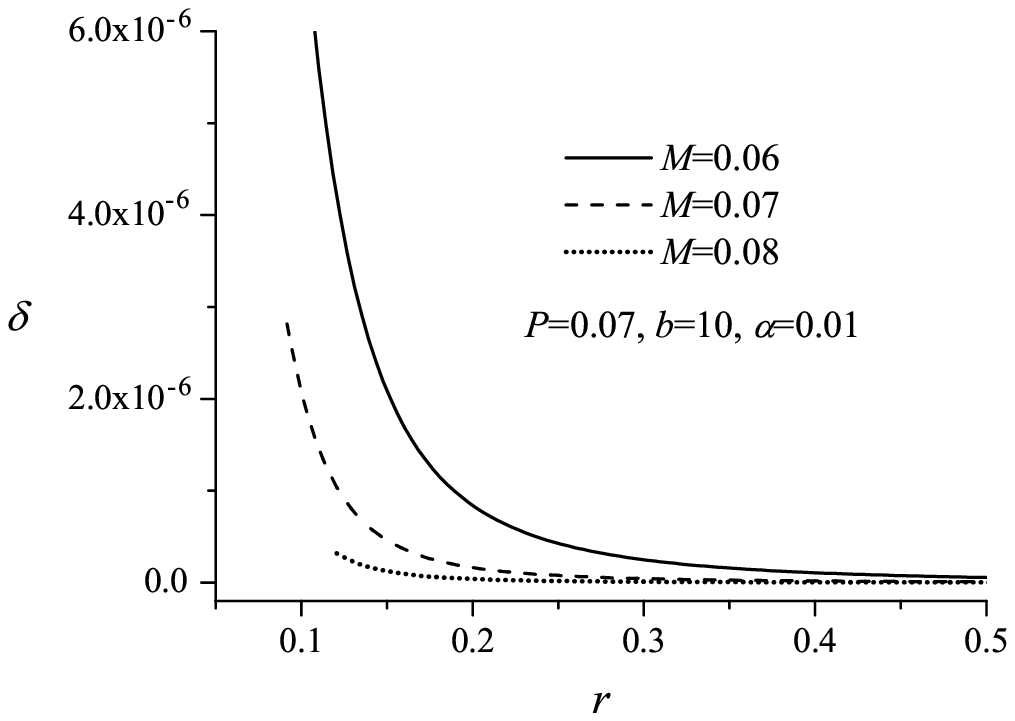}
\includegraphics[width=0.47\textwidth]{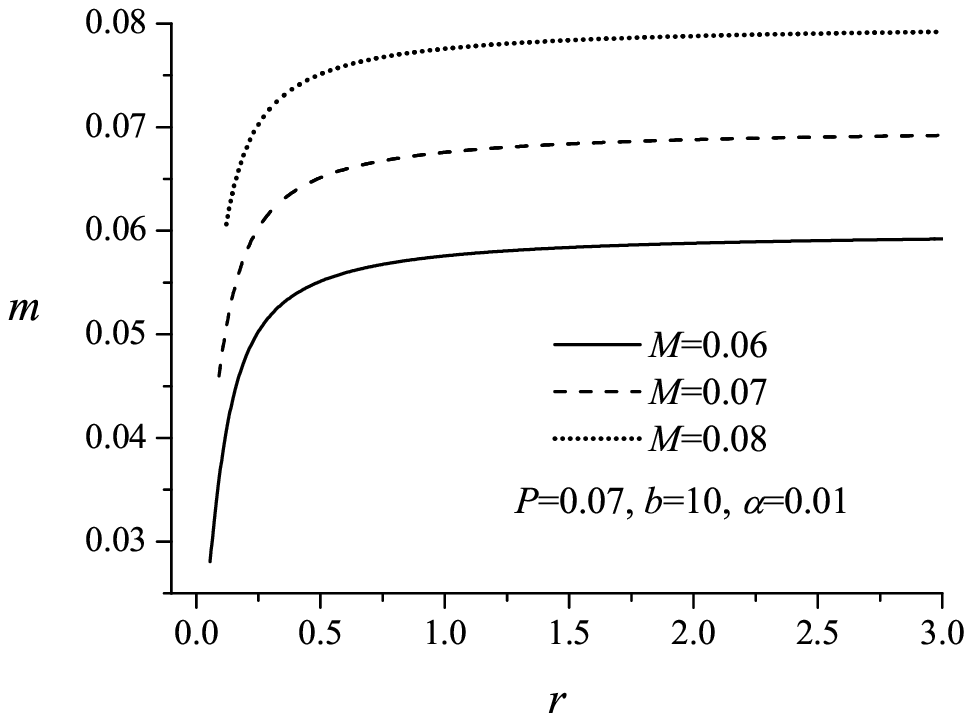}
\caption{%
The metric functions $\delta(r)$ (left panel) and $m(r)$ (right panel) for several values of the black hole mass and for $P=0.07$, $b=10$, and $\alpha=0.01$.} \label{fig:delta(r)_m(r)_P2}%
\end{figure}

\begin{figure}[htbp]%
\vbox{ \hfil \scalebox{0.75}{ {\includegraphics{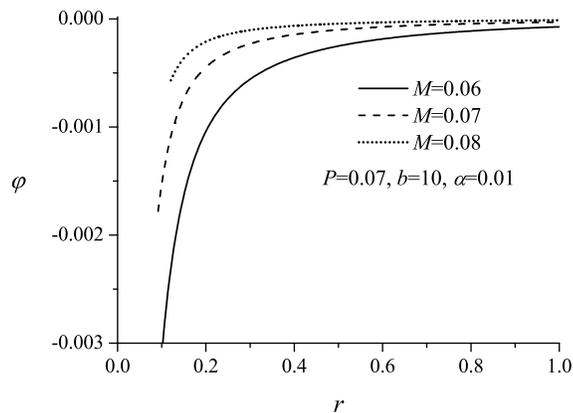}} }\hfil}%
%\bigskip%
\caption{%
%------------------------------
The scalar field $\varphi(r)$ for the same values of the parameters as in Fig.~\ref{fig:delta(r)_m(r)_P2}. }
%------------------------------
\label{fig:phi(r)_P2}%
\end{figure}

In Figs.~\ref{fig:delta(r)_m(r)_P2} and \ref{fig:phi(r)_P2}~~the metric functions $m(r)$ and $\delta(r)$ and the
scalar field $\varphi(r)$ for $P=0.07$, $b=10$, $\alpha=0.01$ and for several values of the black hole mass are presented.
An important justification of our results is that, up to the leading terms, the asymptotic form of the numerically obtained functions
$\delta(r)$, $m(r)$, and $\varphi(r)$  at infinity is the same as what is predicted in (\ref{delta_asymptotic}), (\ref{m_asymptotic}) and
(\ref{phi_asymptotic}), respectively.

\begin{figure}[htbp]%
\includegraphics[width=0.49\textwidth]{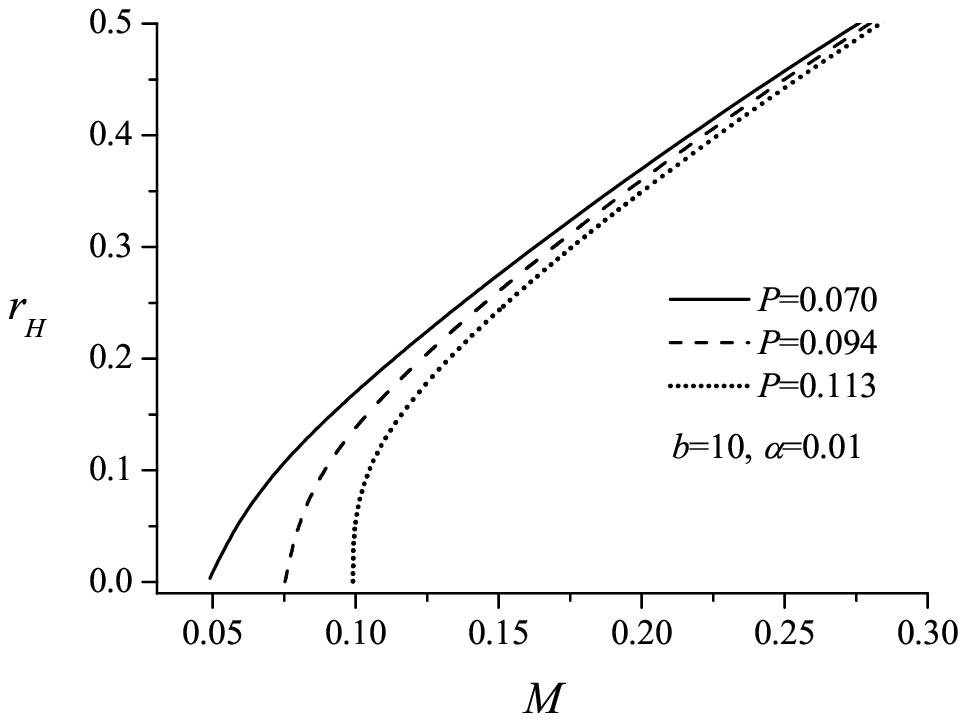}
\includegraphics[width=0.47\textwidth]{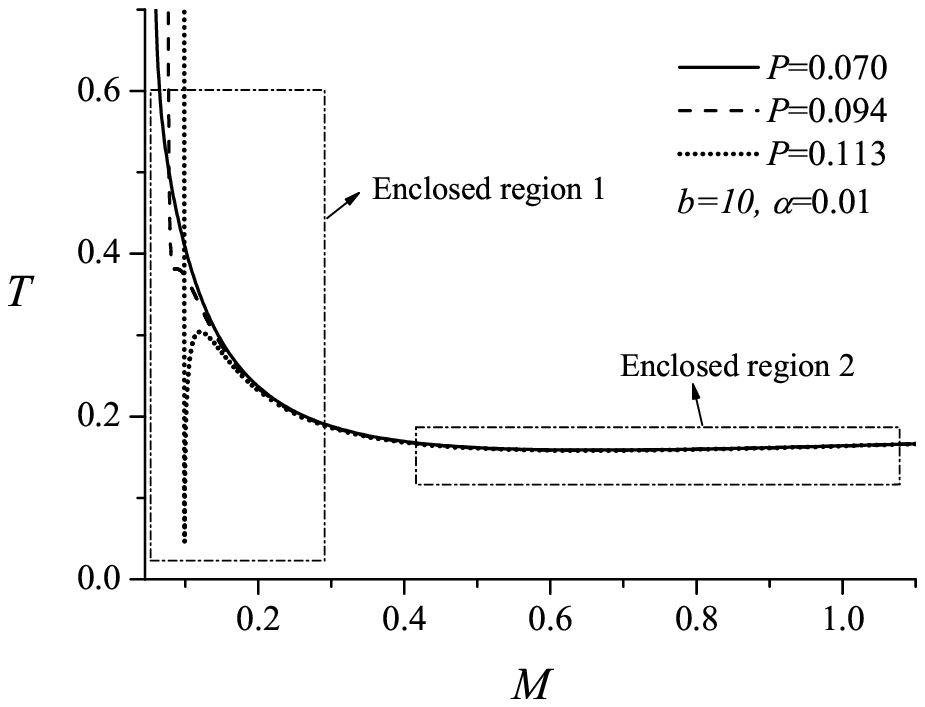}
\caption{%
The radius of the black hole horizon $r_H$ (left panel) and the temperature $T$ (right panel) as functions of the mass $M$ for
sequences of black hole solutions ($b=10$ and $\alpha=0.01$).
Note that on the right panel we denoted ``Enclosed region 1'' and ``Enclosed region 2'' which are expanded in Fig. \ref{fig:T(M)_magnif}.} \label{fig:rh(M)_T(M)}%
\end{figure}

\begin{figure}[htbp]%
\includegraphics[width=0.50\textwidth]{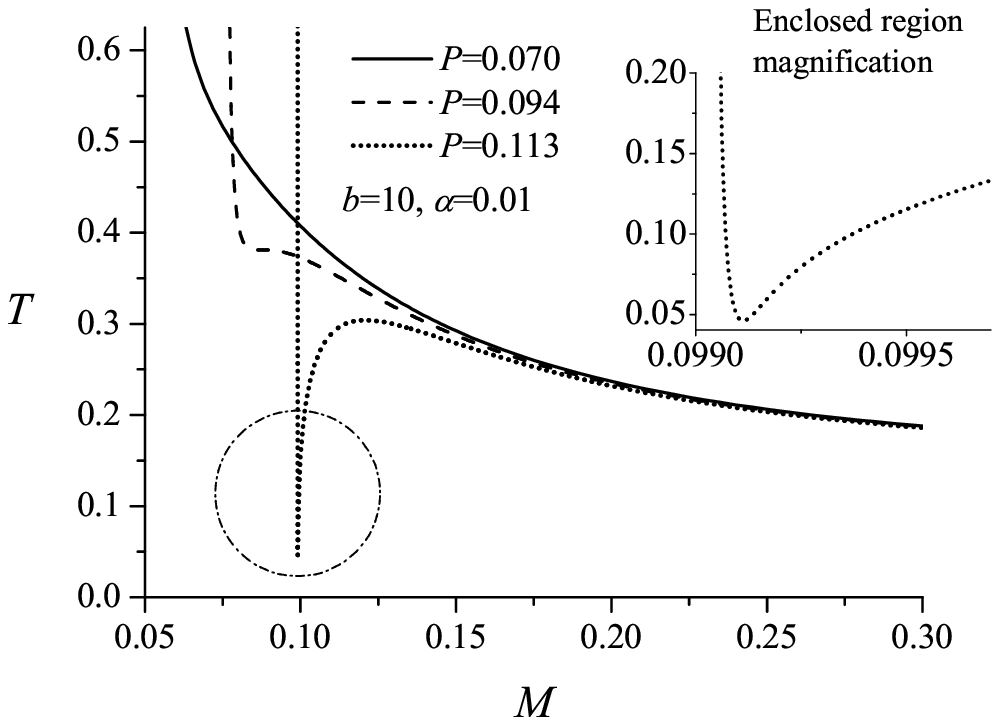}
\includegraphics[width=0.47\textwidth]{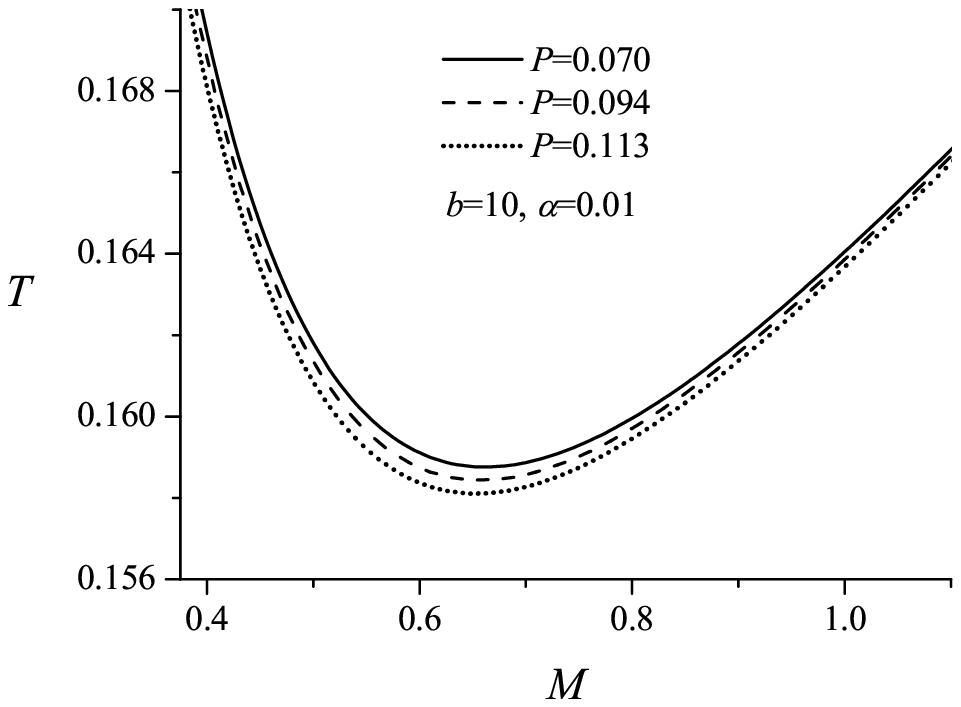}
\caption{%
Magnifications of ``Enclosed region 1'' (left panel) and ``Enclosed region 2'' (right panel) of the $T(M)$ plot in Fig.~\ref{fig:rh(M)_T(M)}.} \label{fig:T(M)_magnif}%
\end{figure}

\begin{figure}[htbp]%
\includegraphics[width=0.47\textwidth]{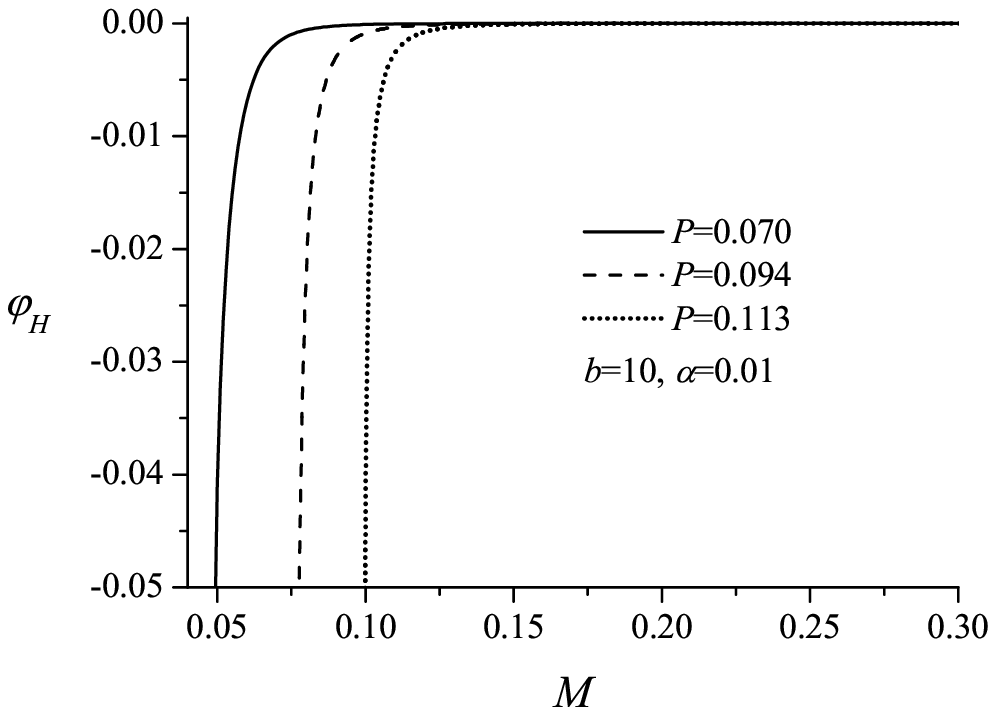}
\includegraphics[width=0.49\textwidth]{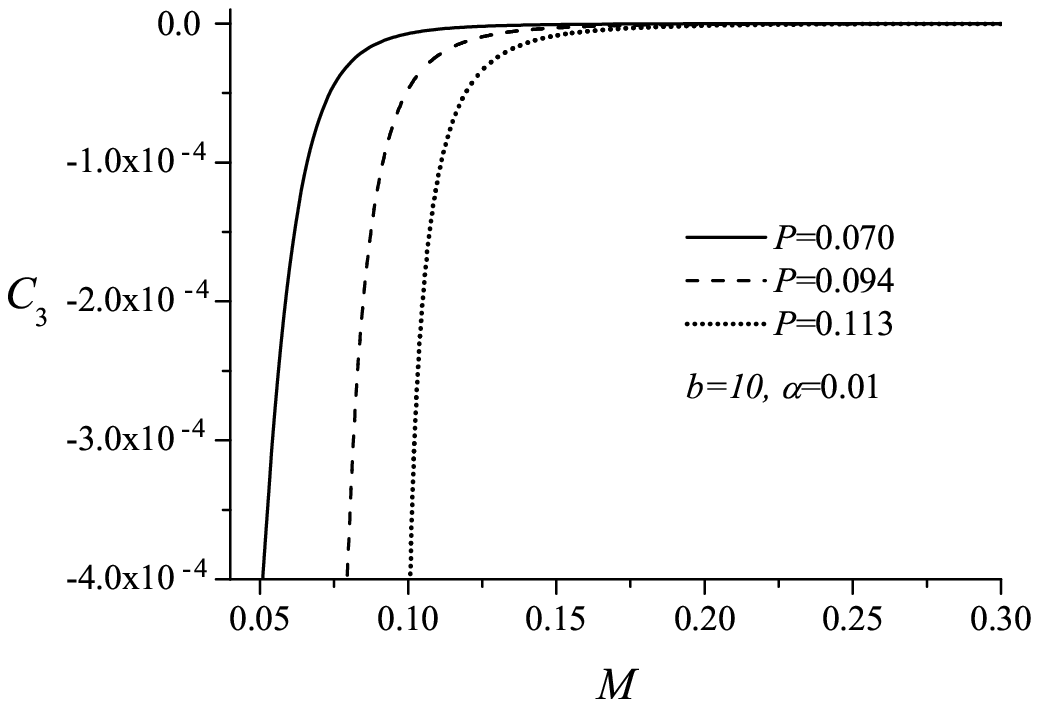}
\caption{%
The scalar field on the horizon $\varphi_H$ (left panel) and the constant $C_3$ (right panel) in the asymptotic expansion of the function $\varphi(r)$
at infinity (see equation (\ref{phi_asymptotic})), as functions of the mass $M$ for
sequences of black hole solutions ($b=10$ and $\alpha=0.01$). } \label{fig:phi_rh(M)_C3(M)}%
\end{figure}

\begin{figure}[htbp]%
\includegraphics[width=0.51\textwidth]{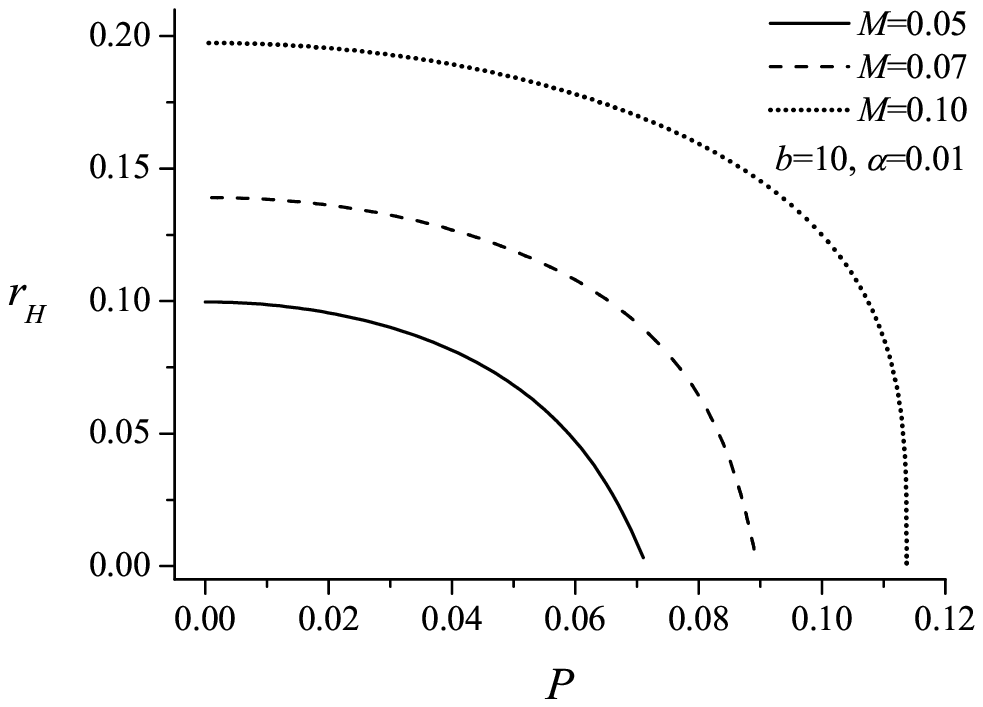}
\includegraphics[width=0.47\textwidth]{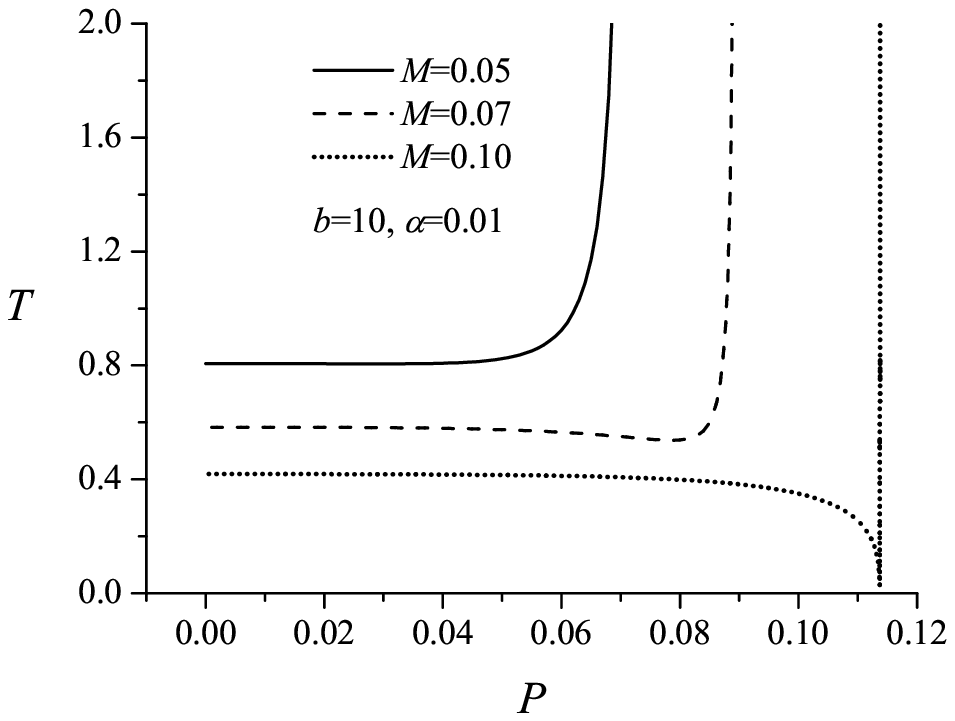}
\caption{%
The radius of the black hole horizon $r_H$ (left panel) and the temperature $T$ (right panel) as functions of the charge $P$ for
sequences of black hole solutions ($b=10$ and $\alpha=0.01$).} \label{fig:RhT(P)}%
\end{figure}

\begin{figure}[htbp]%
\includegraphics[width=0.47\textwidth]{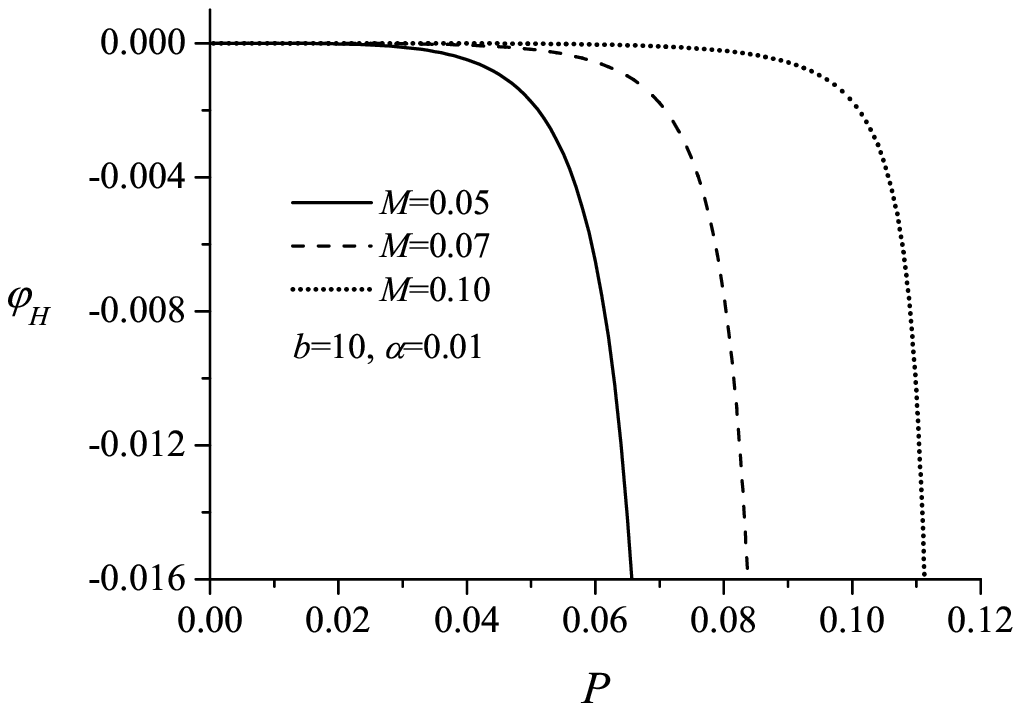}
\includegraphics[width=0.47\textwidth]{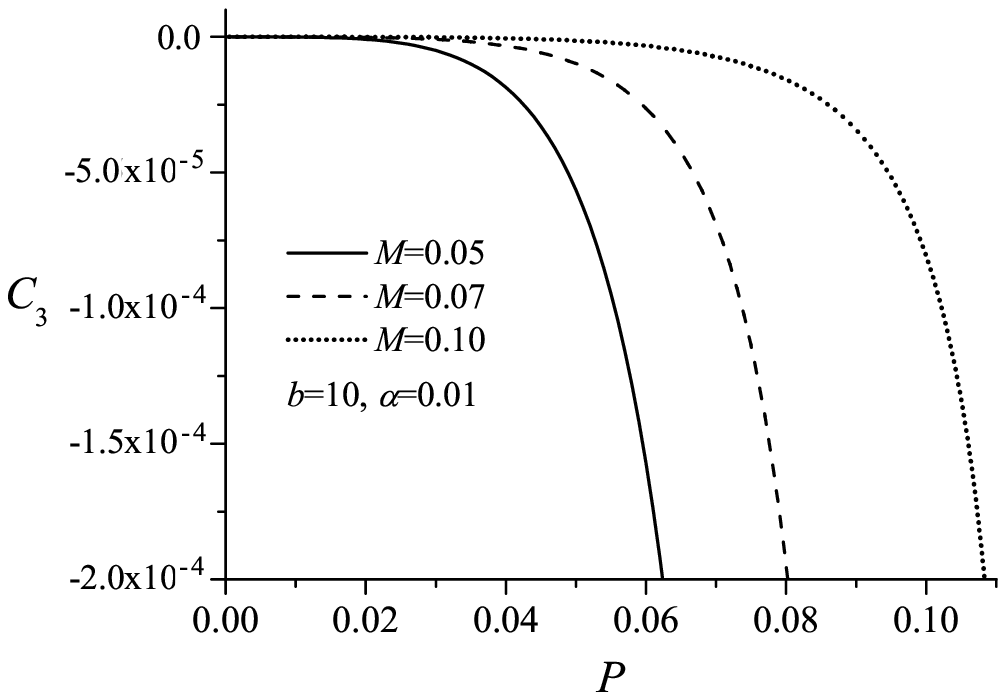}
\caption{%
The scalar field on the horizon $\varphi_H$ (left panel) and the constant $C_3$ (right panel) in the asymptotic expansion of the function $\varphi(r)$ at
infinity (see equation (\ref{phi_asymptotic})), as functions of the charge $P$ for
sequences of black hole solutions ($b=10$ and $\alpha=0.01$).} \label{fig:PhiC3(P)}%
\end{figure}

In Fig.~\ref{fig:rh(M)_T(M)} the horizon $r_H$ and the temperature $T$ as functions of the mass $M$ for sequences of black hole solutions are plotted
for $b=10$ and $\alpha=0.01$. Fig.~\ref{fig:T(M)_magnif} represents two magnifications of the enclosed regions on the $T(M)$ plot in Fig.
\ref{fig:rh(M)_T(M)}. In Fig.~\ref{fig:phi_rh(M)_C3(M)} the value of the scalar field on the event horizon $\varphi_{H}$ and the constant $C_3$
which occurs in the asymptotic expansion of the function $\varphi(r)$ at infinity (see equation (\ref{phi_asymptotic}))  as functions of the mass are
plotted. In Figs.~\ref{fig:RhT(P)} and \ref{fig:PhiC3(P)} the quantities $r_H$, $T$, $\varphi_{H}$ and $C_3$ as functions of the charge $P$ are
plotted.

When $bP^2>1/8$  extremal black hole solutions exist in Einstein-Born-Infeld theory in AdS spacetime. In the presence of scalar field
extremal black holes cannot exist as we have already proved. As we see from the $T(M)$ plot for $P=0.113$ and $b=10$ (this combination satisfies
$bP^2>1/8$) in the right panel of Fig.~\ref{fig:rh(M)_T(M)}, when $M$ decreases the temperature first reaches values close to zero and
afterward increases. The reason is that the absolute value of the scalar field rises considerably in this region, as we can see in
Fig.~\ref{fig:phi_rh(M)_C3(M)}, and prevents the reaching of an extremal black hole. The same situation can be seen from the plot $T(P)$ in Fig.
\ref{fig:RhT(P)} when we increase the magnetic charge $P$ (see, for example, the curve corresponding to $M=0.10$).

The numerical results are obviously subject to and possess the properties in Subsection \ref{3.3}.
All of the dependences have similar qualitative behavior in the parameter space that we have
studied. The only exception is the function $T(M)$ because its qualitative behavior depends also on the value of the parameter $b$ (more details on
this subject will be given in the next sections).

 \pagebreak

\section{Thermodynamic  phase structure }\label{sect_TD}

To study the thermodynamics\footnote{Some clarifying comments on the black hole thermodynamics in scalar-tensor
theories are given in Appendix \ref{STTERMODYNAMICS}. } of the black holes under consideration we need to find the conserved charges of our system first.
Throughout the years many
methods have been developed to calculate the conserved charges of the gravitational configurations. A related problem
is the computation of the gravitational action of a noncompact spacetime. When evaluated on noncompact solutions, the
bulk Einstein term and the boundary Gibbons-Hawking term are both divergent. The remedy is to consider these quantities
relative to those associated with some background reference spacetime, whose boundary at infinity has the same induced
metric as that of the original spacetime. This substraction procedure is however connected to difficulties: the choice
of the reference background is by no means unique and it is not always possible to embed a boundary with a given
induced metric into the reference background. A new method, free of the aforementioned difficulties, was proposed in
\cite{BKraus}. This method (called the counterterm method) consists of adding a (counter)term to the boundary at infinity,
which is a functional only of the curvature invariants of the induced metric on the boundary. Unlike the substraction
procedure, this method is intrinsic to the spacetime of interest and is unambiguous once the counterterm which cancels
the divergences is specified. In the present work we use, namely, the counterterm method in order to compute the mass
and gravitational action. The technical details are presented in Appendix \ref{TD_action}.

We will study the thermodynamic phase structure for  black holes in the  canonical
ensemble where the magnetic  charge $P$ is fixed. The associated thermodynamical potential in the case of the canonical ensemble, as one
should expect from pure thermodynamical considerations, is the Helmholtz free energy $F(T,P)=M-TS$. This can be proven with more rigor by
calculating the Euclidean
action as it is done in Appendix \ref{TD_action} .

Let us note that our definition of the free energy is intrinsic. Our computation based on the counterterm method  makes no reference to any
other solution of the field equations. This should be contrasted with the method used in \cite{Myung3} where
the authors  compute the action using the extremal solution (with fixed charge) as reference background. As we already showed in
our case there are no extremal solutions and therefore the reference background method of  \cite{Myung3} is unapplicable.

The thermodynamic phase structure depends on the value of magnetic charge $P$ and the parameter $b$ in the
Born-Infeld Lagrangian. The numerical results show that for large values of the parameter $b$ the function $T=T(r_H,P)$
has two inflection points defined by
\begin{eqnarray}
{\partial T\over \partial r_H}={\partial^2 T\over \partial r^2_H}=0 . \label{inflections}
\end{eqnarray}
These inflection points, which we will denote by $P^{(1)}_{\rm crit}$ and $P^{(2)}_{\rm crit}$, separate three charge
intervals given by $P<P^{(1)}_{\rm crit}$, $P^{(1)}_{\rm crit}<P<P^{(2)}_{\rm crit}$ and $P^{(2)}_{\rm crit}<P$. Below we will consider
each of these intervals in detail.

For small values of the parameter $b$ the inflection points disappear and the phase structure is the same for
all the charges. The numerical results show that for all of the studied values of the parameter $\alpha$ (including the
case $\alpha=0$ which corresponds to pure Einstein gravity) the critical value of the parameter $b$ which separates the
regions of small and large $b$, is $b_{\rm crit}\approx0.6$.

\subsection{Phase structure for large values of the parameter $b>b_{\rm crit}$}
As we noted for large values of the parameter $b>b_{\rm crit}$ the function $T=T(r_H,P)$ has two inflection points
-- $P^{(1)}_{\rm crit}$ and $P^{(2)}_{\rm crit}$, which gives us three charge intervals defined by $P<P^{(1)}_{\rm crit}$,
$P^{(1)}_{\rm crit}<P<P^{(2)}_{\rm crit}$ and $P^{(2)}_{\rm crit}<P$. Formally we call these three cases black hole
with small, intermediate and large charge respectively. On its own side the intermediate charge interval has to be divided into three
subintervals $P^{(1)}_{\rm crit}<P<P^{(1)}_{\rm ph}$, $P^{(1)}_{\rm ph}<P<P^{(2)}_{\rm ph}$ and $P^{(2)}_{\rm ph}<P<P^{(2)}_{\rm crit}$
($P^{(1)}_{\rm ph}$ and $P^{(2)}_{\rm ph}$ are defined and discussed in details in the following subsections).
As we show below, there are phase transitions for charges belonging to the second and the third subintervals.

All the results for the scalar-tensor black holes given in this subsection are computed for fixed values of the parameters $b=10$ and $\alpha=0.01$. For this choice of parameters the values of the critical charges are  $P^{(1)}_{\rm crit}=0.09397$
and $P^{(2)}_{\rm crit}=0.2901$. The value of $P^{(2)}_{\rm crit}$ is the same (within the numerical error) as in the pure Einstein theory
(i.e. for $\alpha=0$). The reason is that the
corresponding inflection point, defined by equation (\ref{inflections}), is situated in a region of the parameter space where the contribution of the scalar field  is very small.
The value of $P^{(1)}_{\rm crit}$ in the case of pure Einstein theory -- $P^{(1)}_{\rm crit \mbox{ } \alpha=0}=0.09395$,
 is a bit smaller than in the case when scalar field is present.
The numerical results for larger $\alpha$ confirm that $P^{(1)}_{\rm crit}$ increases slightly when we increase the value of the
coupling constant $\alpha$.

The values of the charges $P^{(1)}_{\rm ph}$ and $P^{(2)}_{\rm ph}$ for the chosen values of the parameters $b=10$ and $\alpha=0.01$, are
$P^{(1)}_{\rm ph}=0.1087$ and $P^{(2)}_{\rm ph}=0.1093$ and the corresponding values in pure Einstein theory are $P^{(1)}_{\rm ph \mbox{ } \alpha=0}=0.1085$
and $P^{(2)}_{\rm ph \mbox{ } \alpha=0}=0.1091$. The numerical experiments show that the values of $P^{(1)}_{\rm ph}$ and $P^{(2)}_{\rm ph}$
increase slightly with the increase of $\alpha$.

\begin{figure}[t]%
\vbox{ \hfil \scalebox{1.00}{ {\includegraphics{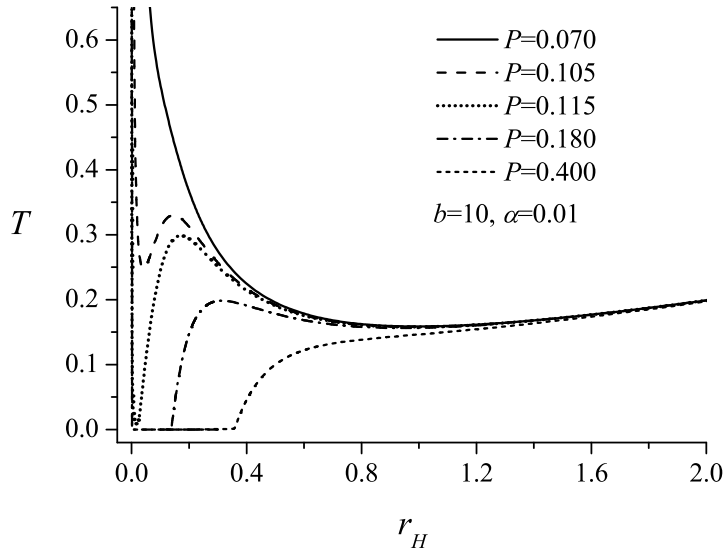}} }\hfil}%
\caption{The temperature $T$ as a function of the radius of the horizon $r_H$ for sequences of black hole solutions ($b=10$ and $\alpha=0.01$).} \label{T_all}
\end{figure}%
\begin{figure}[htbp]%
\vbox{ \hfil \scalebox{1.00}{ {\includegraphics{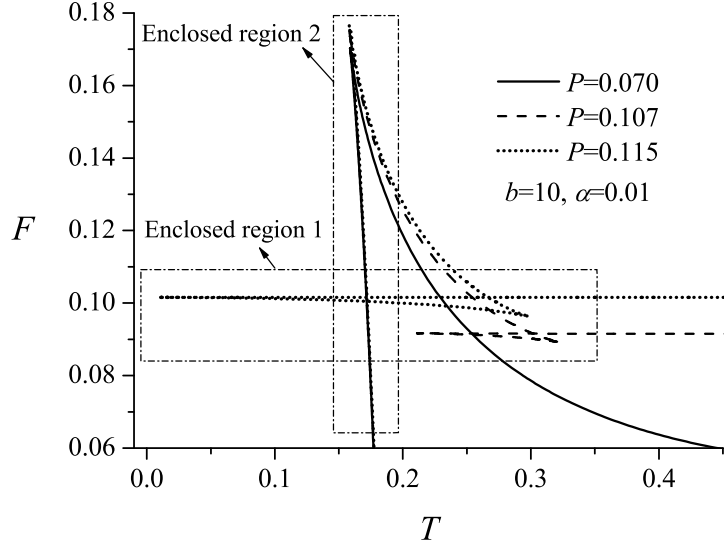}} }\hfil}%
\caption{The free energy $F$ as a function of the temperature $T$ for sequences of black hole solutions
($b=10$ and $\alpha=0.01$). Note that in the figure we defined ``Enclosed region 1''
 and ``Enclosed region 2'' which are expanded in Fig. \ref{F_all_magnif}} \label{F_all}
\end{figure}%
\begin{figure}[htbp]%
\includegraphics[width=0.47\textwidth]{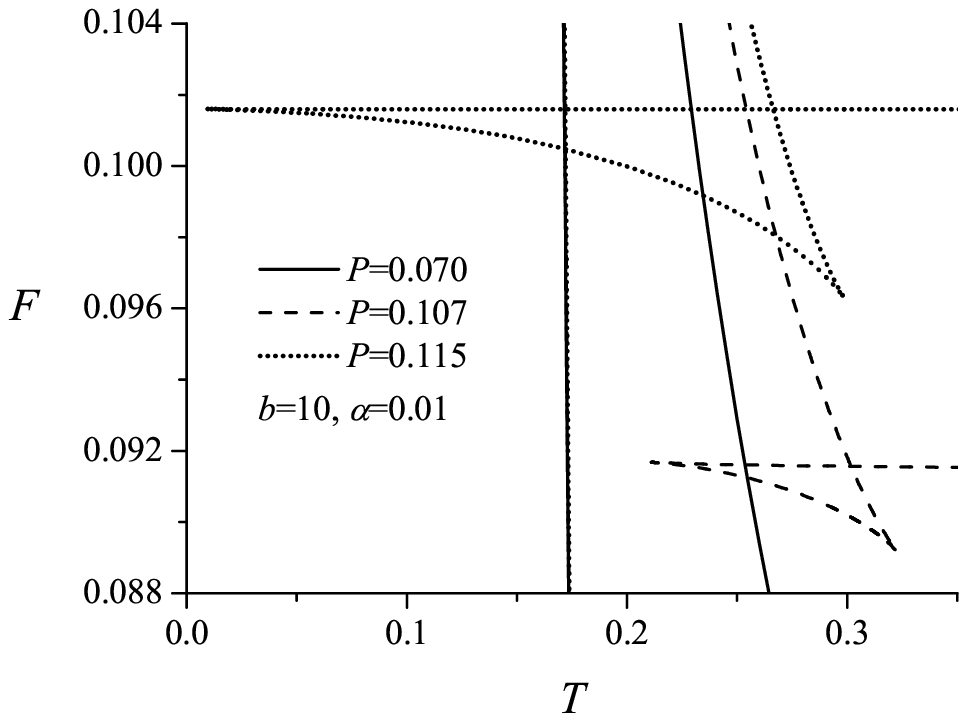}
\includegraphics[width=0.48\textwidth]{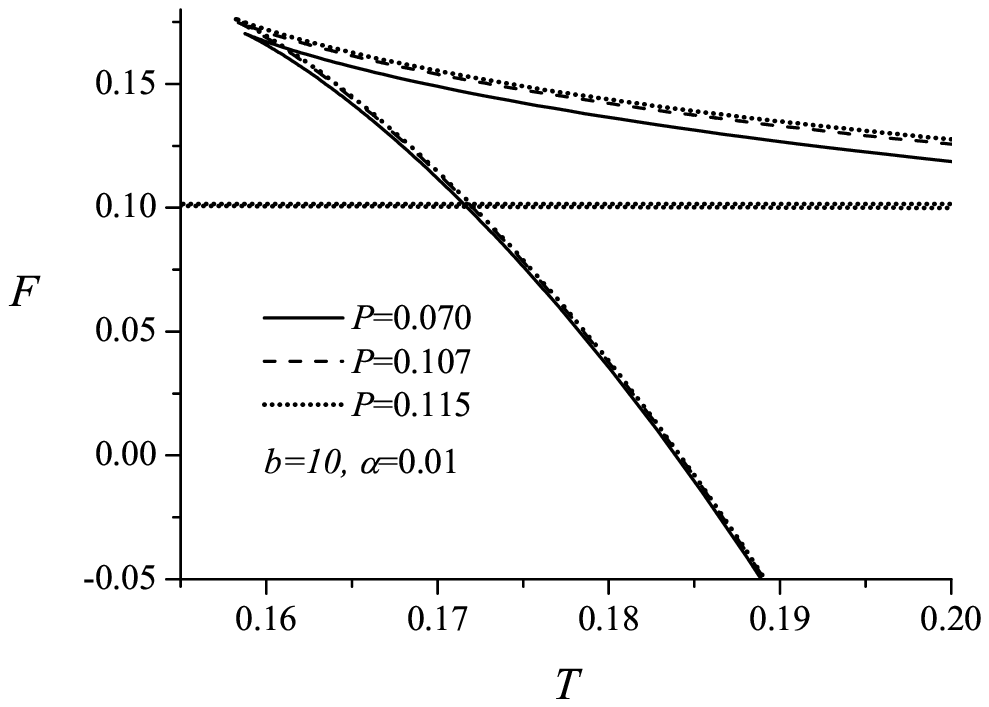}
\caption{Magnifications of the enclosed regions of the $F(T)$ plot in Fig. \ref{F_all}. The left panel corresponds to ``Enclosed
region 1'' and the right panel corresponds to ``Enclosed region 2''.} \label{F_all_magnif}
\end{figure}%
\begin{figure}[htbp]%
\includegraphics[width=0.47\textwidth]{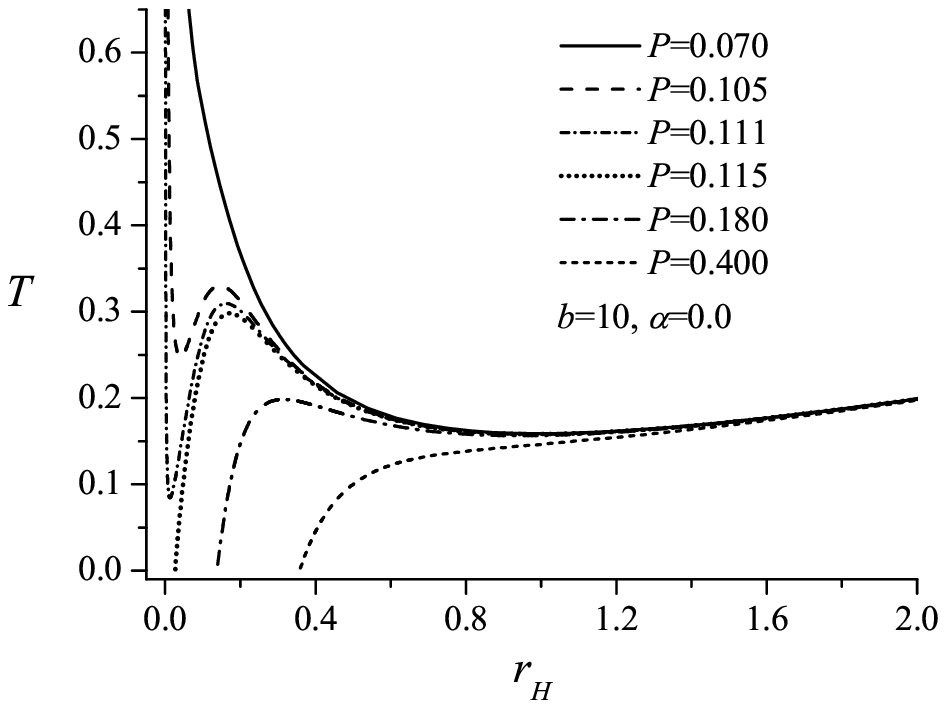}
\includegraphics[width=0.47\textwidth]{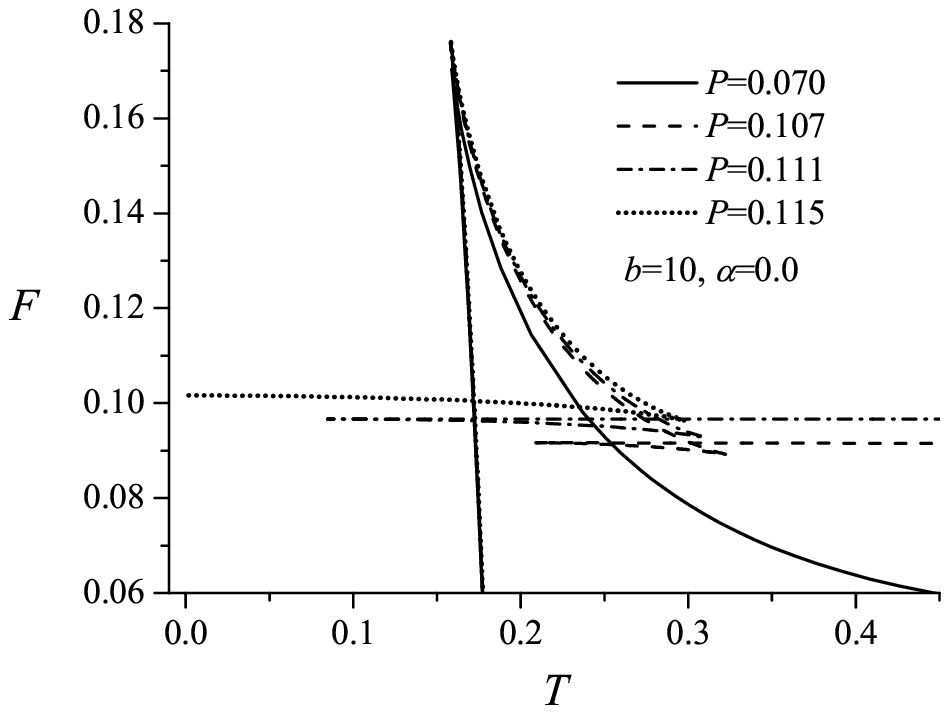}
\caption{The functions $T(r_H)$ and $F(T)$ for sequences of black hole solutions in pure Einstein theory ($b=10$ and $\alpha=0.0$).} \label{FT_all_Alp0}
\end{figure}%
\begin{figure}[htbp]%
\includegraphics[width=0.47\textwidth]{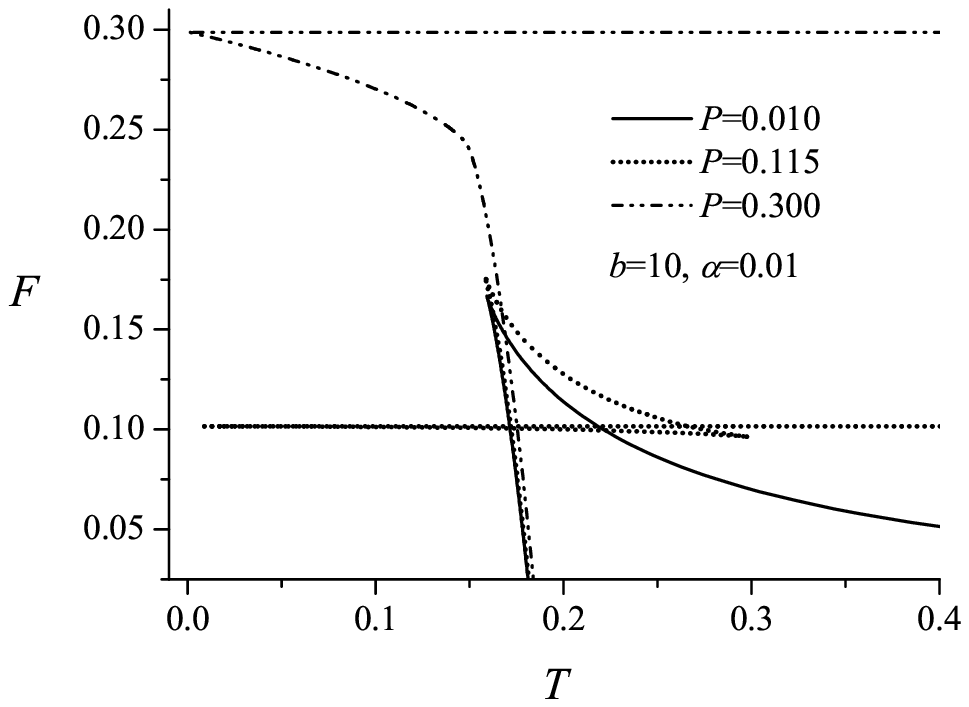}
\includegraphics[width=0.47\textwidth]{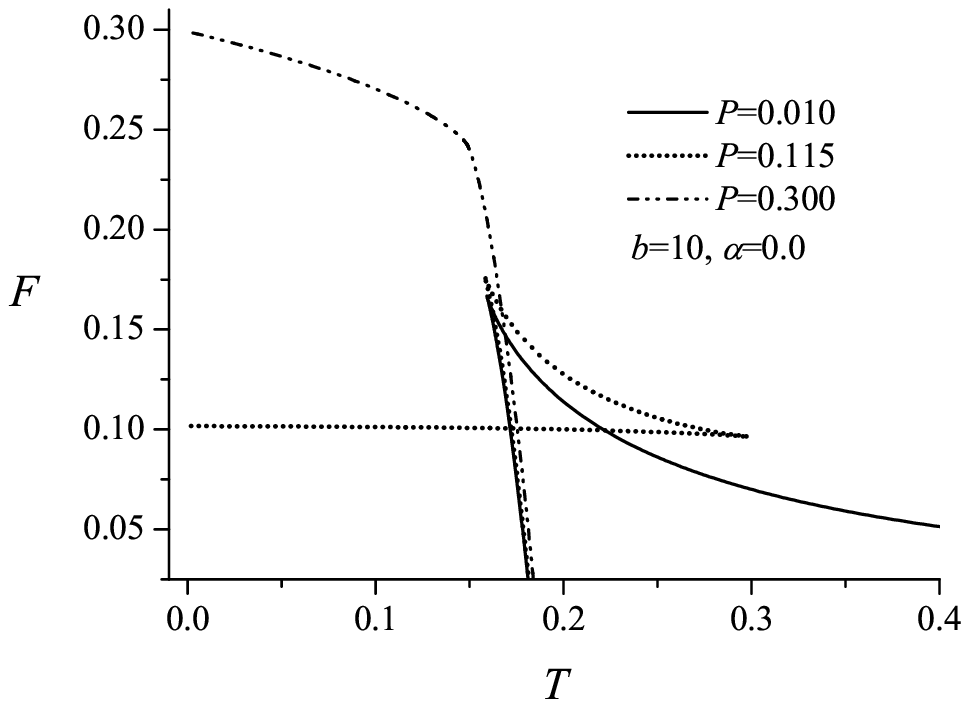}
\caption{The function $F(T)$ for sequences of black hole solutions with various charges $P$, including large charges,
in the case when $\alpha=0.01$ (left panel) and for pure Einstein theory -- $\alpha=0.0$ (right panel).}
\label{F_all_largeP}
\end{figure}%

The temperature $T$ as a function of the horizon radius $r_H$ is shown in Fig.~\ref{T_all} for the cases of small
(for example $P$=0.07), intermediate (for example $P$=0.115), and large (for example $P$=0.40) charges $P$. The free energy
$F$ as a function of the temperature $T$ for small and intermediate charges is shown in Fig.~\ref{F_all} and
magnifications of specific regions are presented in Fig.~\ref{F_all_magnif}. The function $F(T)$ for various charges,
including large charges, is given in the left panel of Fig.~\ref{F_all_largeP}. In Figs.~\ref{FT_all_Alp0} and
\ref{F_all_largeP} (right panel) we also give the
dependences $T(r_H)$ and $F(T)$ for Born-Infeld-AdS (BIAdS) black holes in the pure Einstein gravity for comparison.
The figures show that the thermodynamic phase structures for the scalar-tensor
BIAdS black holes and the Einstein BIAdS black holes are qualitatively the same except in the case of charges satisfying $bP^2>1/8$ where
additional phases of the scalar-tensor BIAdS black holes are present. Below we consider in detail the thermodynamic phase structure for the
intervals of small, intermediate and large charges.

The phase structure and the phase transitions can be alternatively studied with the so-called off-shell formalism or,
in other words, off-equilibrium formalism. It is thoroughly presented in \cite{Arcioni} and applied, for example, in
the papers on black-hole phase transitions by Myung and coauthors \cite{Myung1, Myung2, Myung3}. The advantage of that formalism is that
it gives a simple and clear interpretation of the thermodynamic phases. In the canonical ensemble the different phases of
the system, both stable and unstable, correspond to extrema of the so-called off-shell free energy. In that formalism,
the origin of the branches in the $T-F$ diagram, that correspond to the different phases, can be also simply explained.
The phase transitions have been studied with the application of the off-shell formalism in Appendix \ref{TD_offshell}.

\subsubsection{ Phase structure for small charges}

\begin{figure}[t]%
\includegraphics[width=0.47\textwidth]{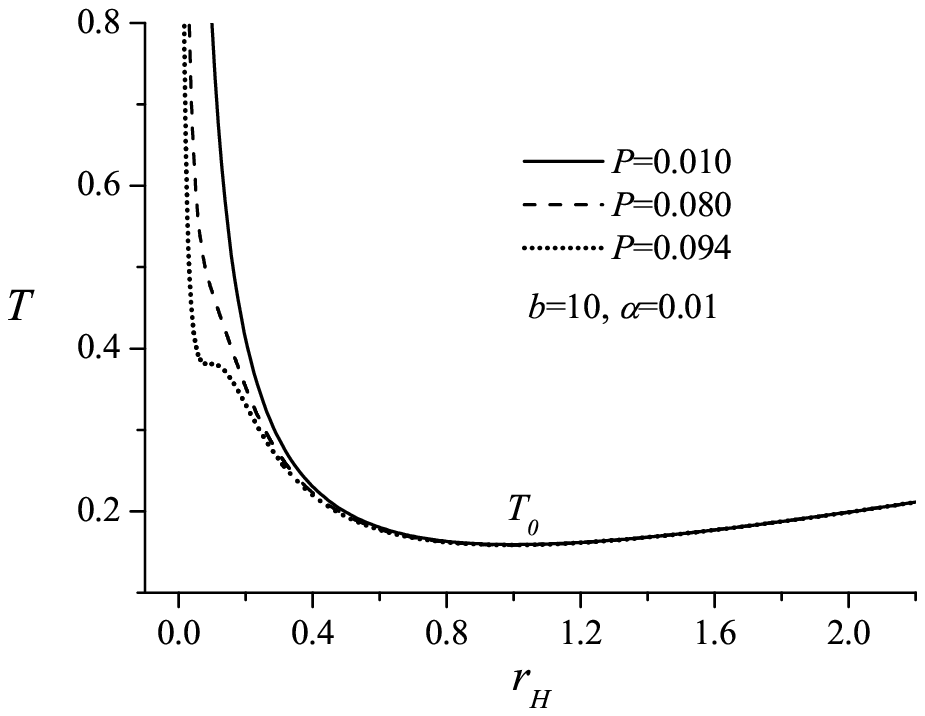}
\includegraphics[width=0.49\textwidth]{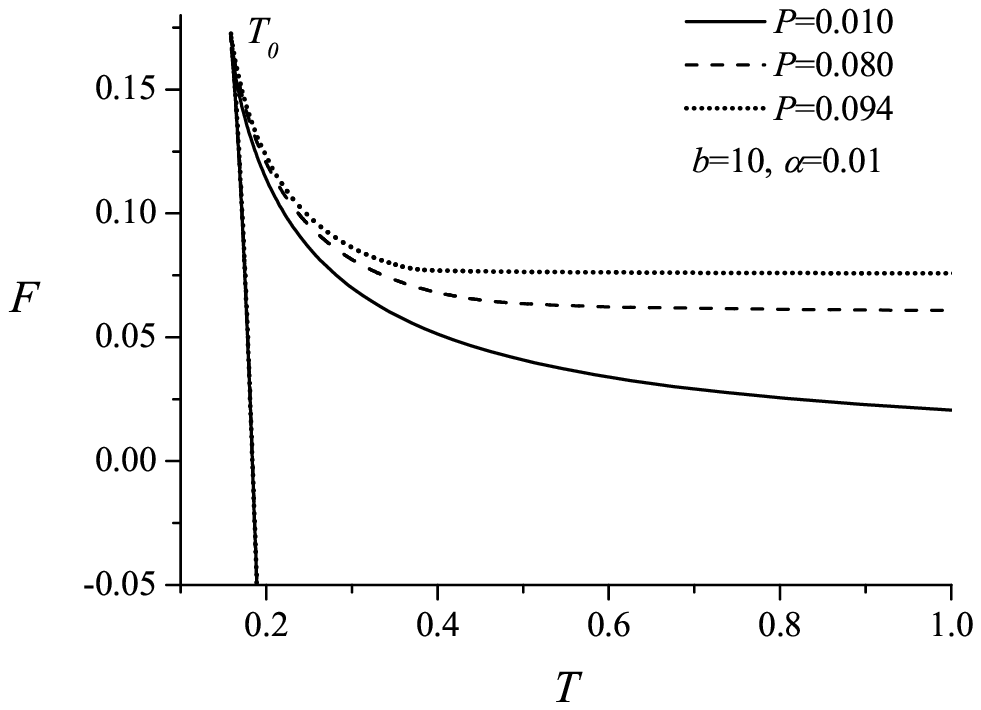}
\caption{The functions $T(r_H)$ and $F(T)$ for small charges where $b=10$ and $\alpha=0.01$.
 The curve corresponding to the critical charge $P^{(1)}_{\rm crit}$ is denoted by the dotted line.} \label{FT_small}
\end{figure}%

In the region of small charges $P<P^{(1)}_{\rm crit}$,
as one can see from Fig. \ref{FT_small}, for arbitrary temperature $T>T_{0}$ there are two branches of solutions for the black hole radius.
These two branches correspond to  small and large black  holes, respectively. The small black holes have negative
specific heat $C_P$ which is defined according to the formula
\begin{eqnarray}
C_P= T \left.{\partial S\over \partial T}\right|_{P}= 2\pi T r_H \left.\left({\partial T\over \partial r_{h}}\right)^{-1}\right|_{P}
\end{eqnarray}
and  therefore they are unstable. The large black holes have positive specific heat and consequently they are locally
stable. Moreover, as can be seen in Fig.~\ref{FT_small} (right panel)  the free energy of the large black holes is
smaller than the free energy of the small black holes and therefore the large black holes are thermodynamically
favorable and dominate the thermodynamic ensemble.

\subsubsection{ Phase structure for intermediate charges}

\begin{figure}[t]%
\vbox{ \hfil \scalebox{0.78}{ {\includegraphics{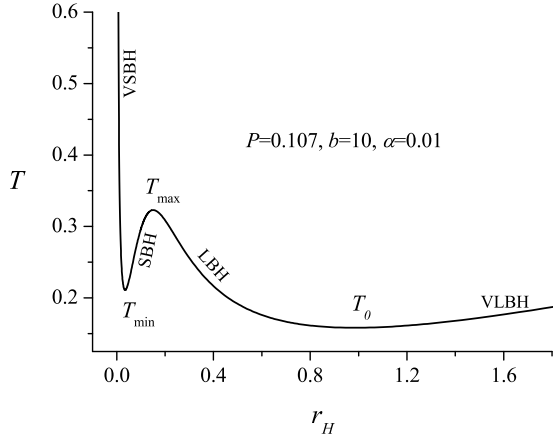}} }\hfil}%
\caption{The function $T(r_H)$ for intermediate charge $P^{(1)}_{\rm crit}<P<P^{(1)}_{\rm ph}$ ($P=0.107$, $b=10$ and $\alpha=0.01$).}\label{T_midd1}
\end{figure}

\begin{figure}[t]%
\includegraphics[width=0.47\textwidth]{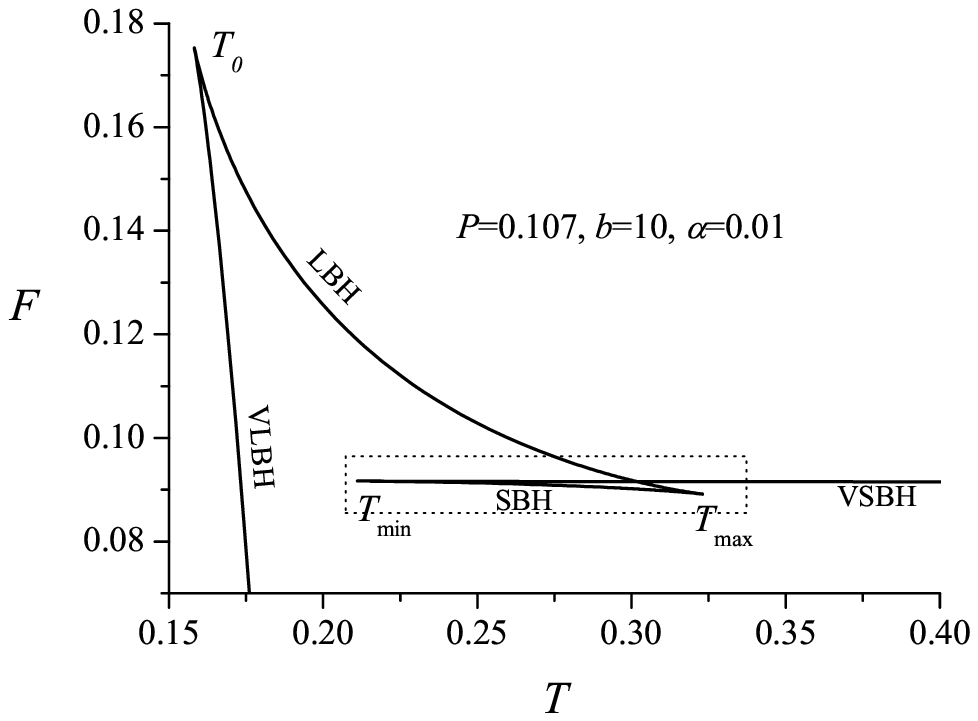}
\includegraphics[width=0.48\textwidth]{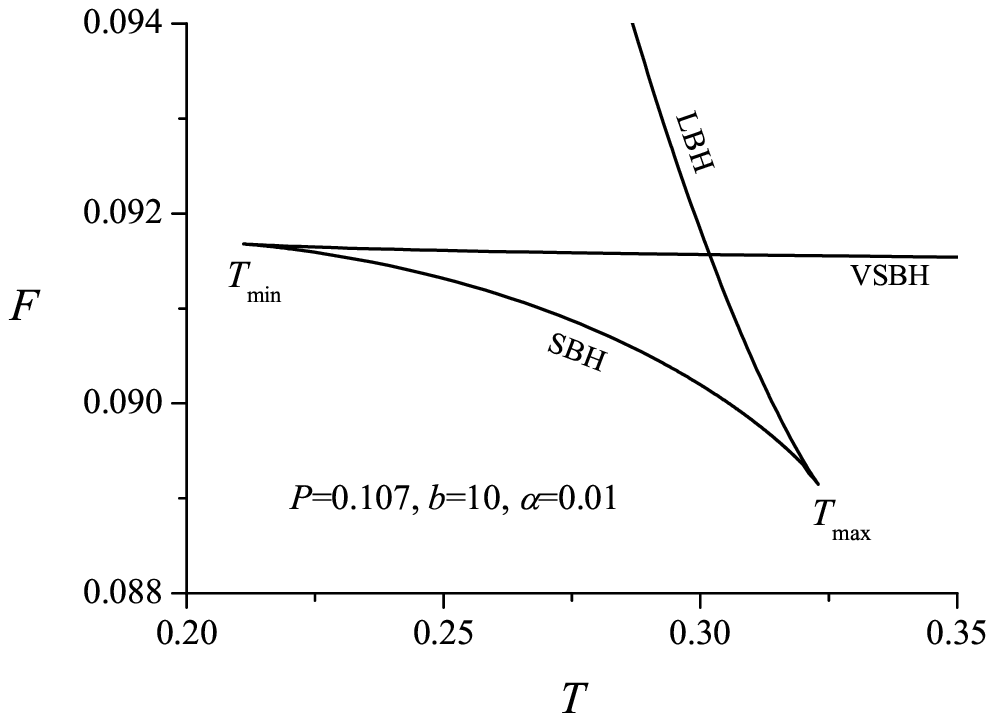}
\caption{The function $F(T)$ for intermediate charge $P^{(1)}_{\rm crit}<P<P^{(1)}_{\rm ph}$ ($P=0.107$, $b=10$ and $\alpha=0.01$). The right
panel is a magnification of the enclosed region of the left figure.} \label{F_midd1}
\end{figure}

\begin{figure}[htbp]%
\includegraphics[width=0.47\textwidth]{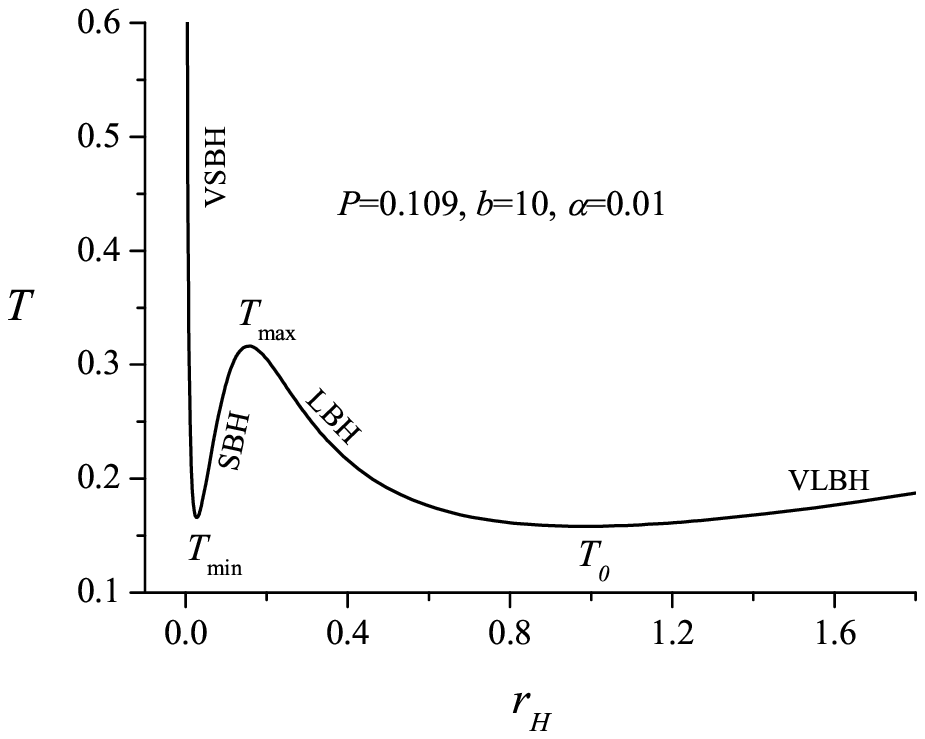}
\includegraphics[width=0.47\textwidth]{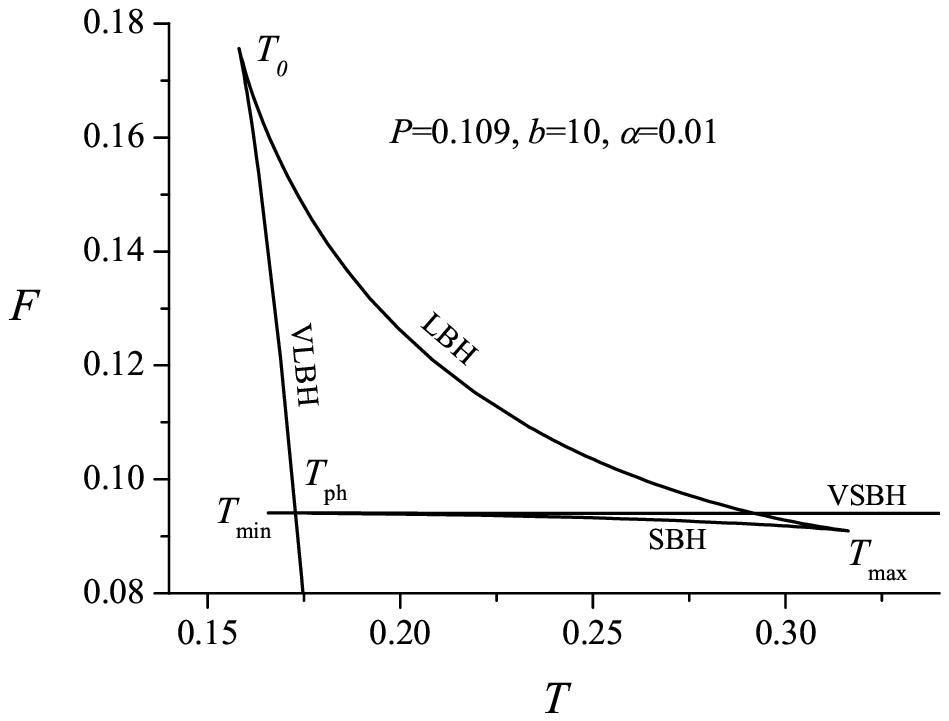}
\caption{The functions $T(r_H)$ and $F(T)$ for intermediate charges $P^{(1)}_{\rm ph}<P<P^{(2)}_{\rm ph}$} ($P=0.109$, $b=10$ and $\alpha=0.01$). \label{FT_interm_2_1}
\end{figure}

\begin{figure}[t]%
\includegraphics[width=0.47\textwidth]{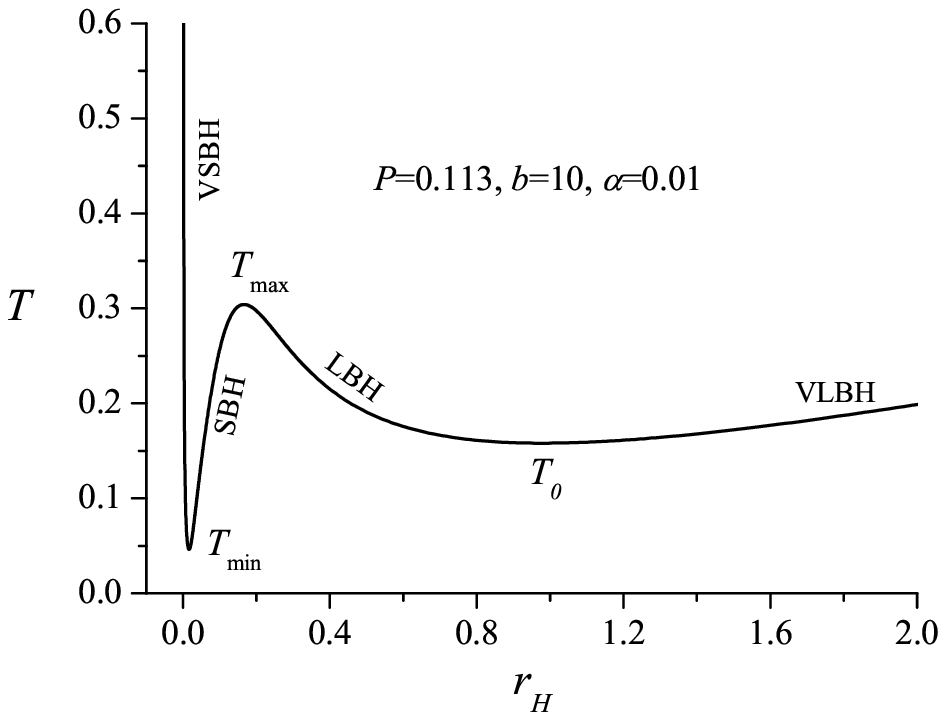}
\includegraphics[width=0.47\textwidth]{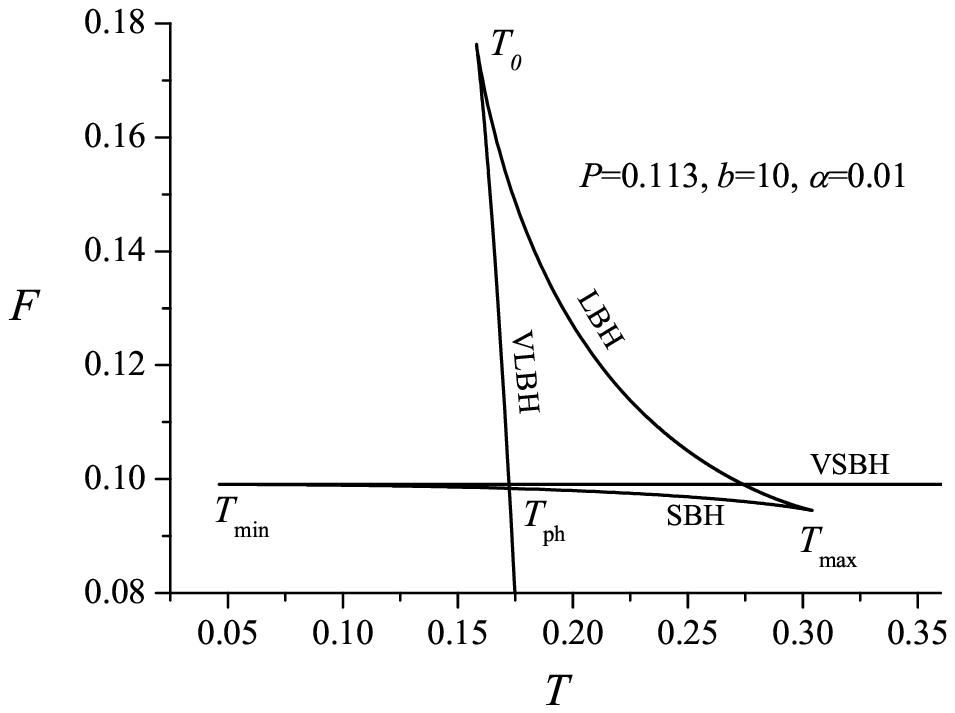}
\caption{The functions $T(r_H)$ and $F(T)$ for intermediate charge $P^{(2)}_{\rm ph}<P<P^{(2)}_{\rm crit}$ ($P=0.113$, $b=10$ and $\alpha=0.01$).}
 \label{FT_interm_2_2} \label{FT_midd2}
\end{figure}

In the intermediate charge interval defined by $P^{(1)}_{\rm crit}<P<P^{(2)}_{\rm crit}$, there are four branches, as shown in
Figs.~\ref{T_midd1}--\ref{FT_interm_2_2}. These branches will be called branches
of very small (VSBH), small (SBH), large (LBH), and very large (VLBH) black holes, respectively.

Below we describe the phase structure in the subintervals $P^{(1)}_{\rm crit}<P<P^{(1)}_{\rm ph}$, $P^{(1)}_{\rm ph}<P<P^{(2)}_{\rm ph}$  and
$P^{(2)}_{\rm ph}<P<P^{(2)}_{\rm crit}$. The differences between the three subintervals are that
in the first subinterval no phase transitions exist, in the second subinterval phase transitions of first and
zeroth order exist, and in the third subinterval only first order phase transition exists. Such subintervals with the corresponding
phase transitions are also present for pure Einstein-Born-Infeld-AdS black holes.
\\
\\
\paragraph{Intermediate charge subinterval $P^{(1)}_{\rm crit}<P<P^{(1)}_{\rm ph}$}
\mbox{ }\\

In this subinterval $T_0<T_{\rm min}$ \footnote{By $T_0$ and $T_{\rm min}$ we denote the two minima of the $T(r_H)$ function as
can be seen in Fig. \ref{T_midd1}.}. The $T(r_H)$ and $F(T)$ dependences are shown in Figs.~\ref{T_midd1} and \ref{F_midd1}
and the phase structure is described as follows.
For low temperature $T$ satisfying $T_0 <T< T_{\rm min}$ there are two
solutions for the black hole horizon radius corresponding to LBH and VLBH. From the dependence $T(r_H)$ shown in Fig.
\ref{T_midd1} one can conclude that the LBH are thermodynamically unstable since they have negative specific heat,
while the VLBH have positive specific heat and therefore they are locally stable.  At $T=T_{\rm min}$ the origin of two new
branches of solutions appears and for $T_{\rm min}<T<T_{\rm max}$ we have four black holes  with the same temperature and
different radii -- VSBH, SBH, LBH, and VLBH. The VSBH and LBH are unstable since their specific heat is
negative, while the SBH and VLBH have positive specific heat and consequently they are locally stable. At $T=T_{\rm max}$
two of the branches coalesce and disappear, leaving only two branches for $T>T_{\rm max}$ corresponding to unstable VSBH
and locally stable VLBH.
From $T=T_0$ to arbitrary large temperature the free energy of the VLBH, corresponding to the
leftmost branch in Fig.~\ref{F_midd1} (left panel) (corresponding to the rightmost branch in Fig.~\ref{T_midd1}), is
the smallest  which means that for arbitrary temperature the VLBH dominate the thermodynamic ensemble.

\paragraph{Intermediate charge subinterval $P^{(1)}_{\rm ph}<P<P^{(2)}_{\rm ph}$}
\mbox{ }\\

In this subinterval $T_0<T_{\rm min}$. The $T(r_H)$ and $F(T)$ dependences are shown in Fig.~\ref{FT_interm_2_1} and the phase structure is
described as follows. For $T_0<T<T_{\rm min}$ we have two branches  -- unstable LBH and locally stable VLBH. At $T=T_{\rm min}$
two new branches appear and for $T_{\rm min}<T<T_{\rm max}$ there are four branches corresponding to  unstable VSBH, locally
stable SBH, unstable LBH and locally stable VLBH. At $T_{\rm max}$ two of the branches coalesce and disappear and we are
left with only two branches of unstable VSBH and locally stable VLBH.  From $T_0$ to $T_{\rm min}$ the VLBH have the
smallest free energy and they dominate the thermodynamical ensemble. At $T=T_{\rm min}$ the situation changes and for temperatures in the
interval from $T_{\rm min}$ to $T_{\rm ph}$ the
ensemble is  dominated by the SBH which have the smallest free energy. At this point $T=T_{\rm min}$ phase transition between VLBH and SBH
occurs and it is zeroth order phase transition since the free energy is discontinuous at that point -- the free energy jumps to a lower value\footnote{We have
studied the zeroth-order phase transition observed at that point with the off-shell formalism in Appendix \ref{TD_offshell}. }. Such unusual phase
transitions of zeroth order have been observed for example in \cite{Vega, Jagla, Maslov}.

As we said for temperatures in the interval $T_{\rm min}<T<T_{\rm ph}$ the
ensemble is  dominated by the SBH. As we can see in Fig.~\ref{FT_interm_2_1} for arbitrary $T>T_{\rm ph}$ the ensemble is
again dominated by the VLBH which have the smallest free energy. This means that at the point $T=T_{\rm ph}$ we observe a phase
transition between the SBH and VLBH and it is classified as a first order phase
transition since the free energy is continuous  at $T=T_{\rm ph}$ and the first derivative of the free energy is discontinuous at that point.

\paragraph{Intermediate charge subinterval $P^{(2)}_{\rm ph}<P<P^{(2)}_{\rm crit}$}
\mbox{ }\\

In this subinterval $T_{\rm min}<T_{0}$. The $T(r_H)$ and $F(T)$ dependences are shown in Fig.~\ref{FT_interm_2_2} and the phase structure is
described as follows. For $T_{\rm min}<T<T_{0}$ there are two branches
corresponding to unstable VSBH and locally stable SBH. At $T=T_{0}$ two new branches appear and for temperatures $T_0<T<T_{\rm max}$ we
have four branches corresponding to unstable VSBH,  locally stable SBH, unstable LBH and locally stable VLBH. At $T=T_{\rm max}$
the branches of SBH and LBH disappear and we are left with unstable VSBH and locally stable VLBH for $T>T_{\rm max}$.
From $T_{\rm min}$ to temperature $T_{\rm ph}$  the free energy of the SBH is the smallest and the SBH dominate the thermodynamic ensemble.
At temperature  $T=T_{\rm ph}$ the free energy of the VLBH becomes smaller than the free energy of the SBH and for all temperatures $T>T_{\rm ph}$ the
VLBH dominate the thermodynamic ensemble. At the point $T=T_{\rm ph}$ we have a first-order phase transition from SBH to VLBH.

\begin{figure}[t]%
\includegraphics[width=0.48\textwidth]{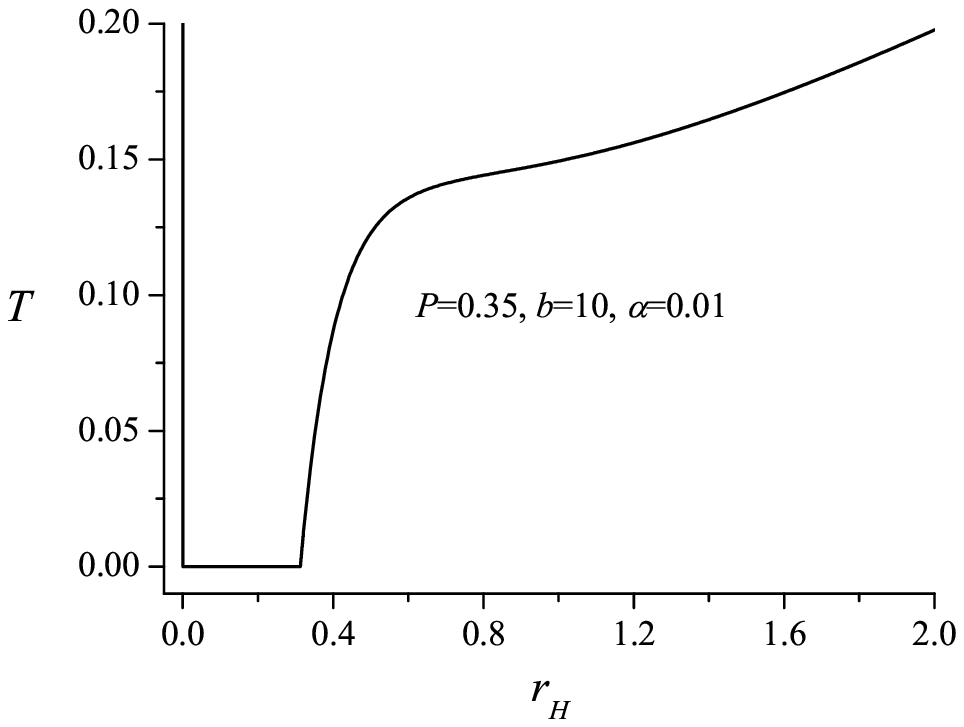}
\includegraphics[width=0.47\textwidth]{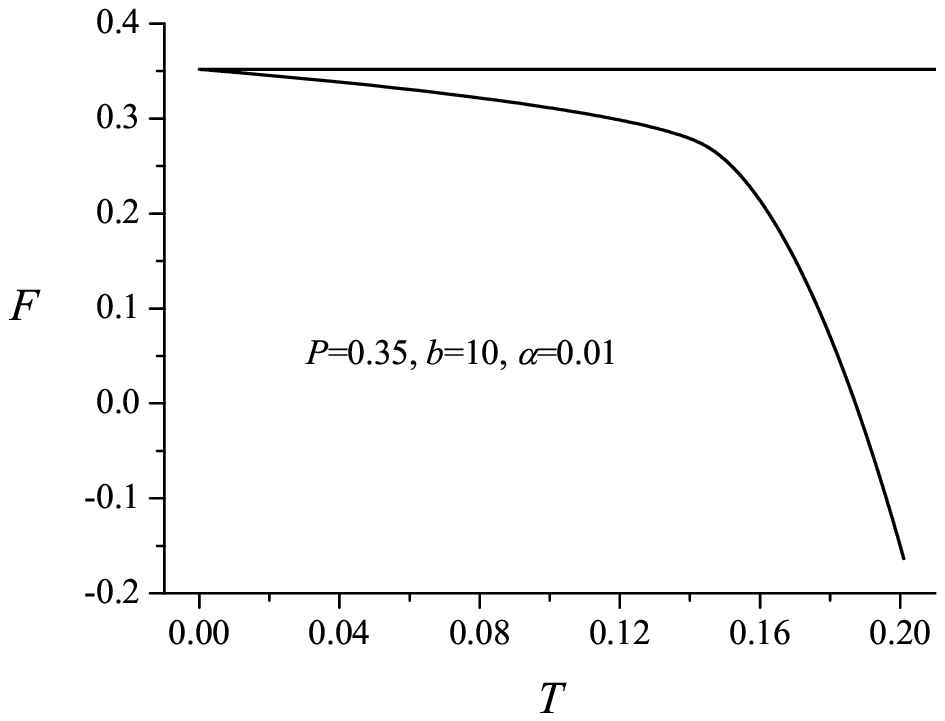}
\caption{The functions $T(r_H)$ and $F(T)$ for large charge $P^{(2)}_{\rm crit}<P$ ($P=0.35$, $b=10$ and
$\alpha=0.01$).}\label{TF_MnGolqmo}
\end{figure}%

\begin{figure}[htbp]%
\includegraphics[width=0.48\textwidth]{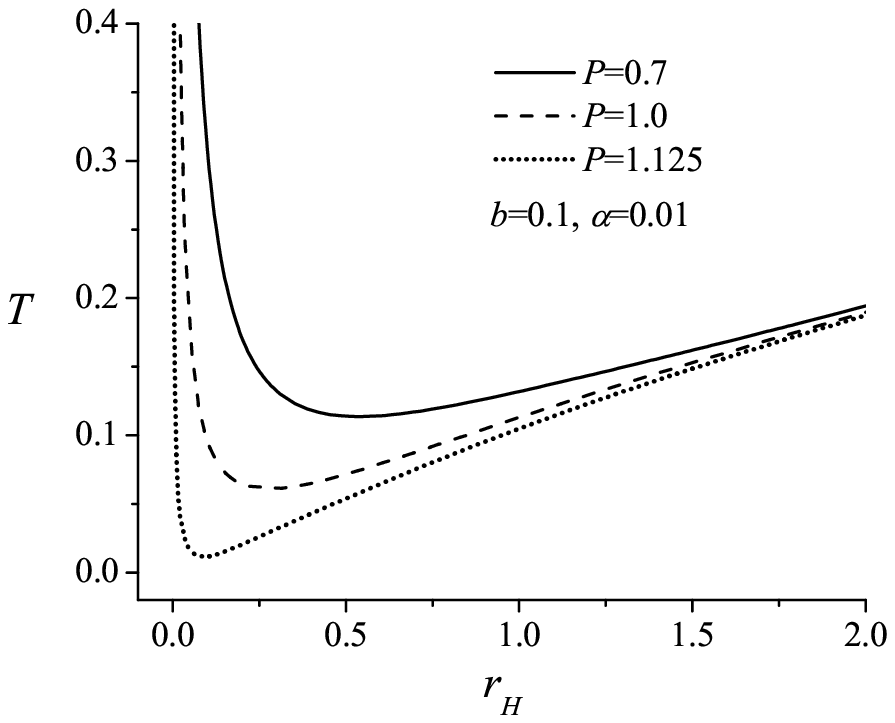}\quad
\includegraphics[width=0.47\textwidth]{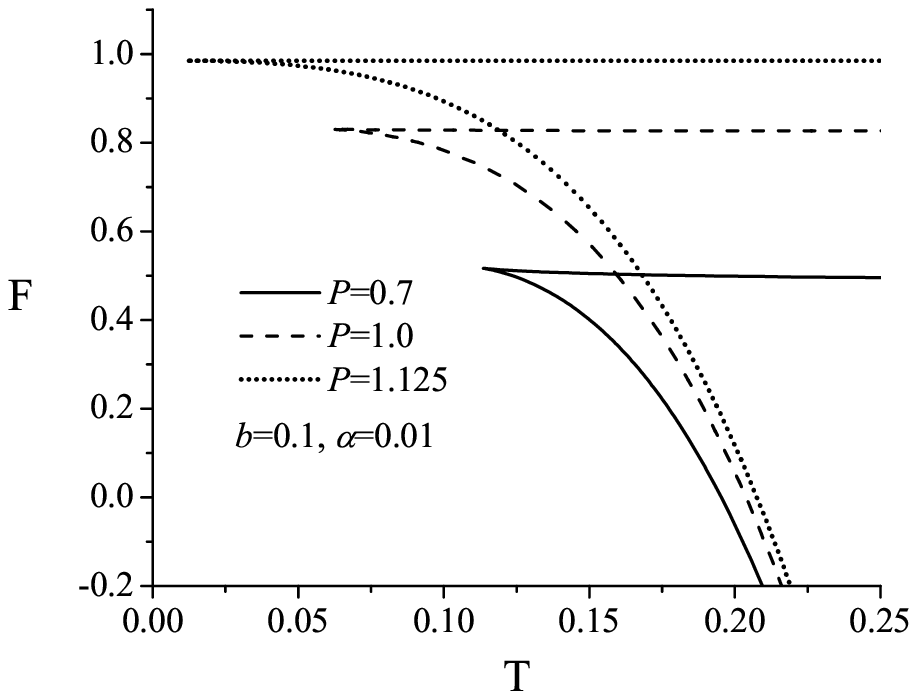}
\caption{The functions $T(r_H)$ and $F(T)$ for a small value of the Born-Infeld parameter $b=0.1$ ($\alpha=0.01$).}\label{TF_bmalko}
\end{figure}%

\subsubsection{ Phase structure for large charges}
In the case of large charges defined by $P^{(2)}_{\rm crit}<P$, there are two branches -- small and large black holes, as Fig.~\ref{TF_MnGolqmo} shows.
The small black holes have negative specific heat and consequently they are  unstable. The large black holes are locally stable since
their specific heat is positive. For arbitrary temperature the free energy of the large black holes is smaller than the free energy of the small
black holes. Therefore the large black holes dominate the thermodynamic ensemble.

\subsection{Phase structure for small values of the parameter $b<b_{\rm crit}$}
The phase structure for small values of the parameter $b$ is much simpler because the function $T=T(r_H,P)$ has
no inflection point. Therefore the phase structure is qualitatively the same for all values of the magnetic charge $P$.
The functions $T(r_H)$ and $F(T)$ are plotted in Fig.~\ref{TF_bmalko} for $b=0.1$ and $\alpha=0.01$. Only two
branches exist corresponding to unstable small black holes and locally stable large black hole. The large
black holes have lower free energy and therefore they dominate the thermodynamic ensemble.

\begin{figure}[t]%
\includegraphics[width=0.48\textwidth]{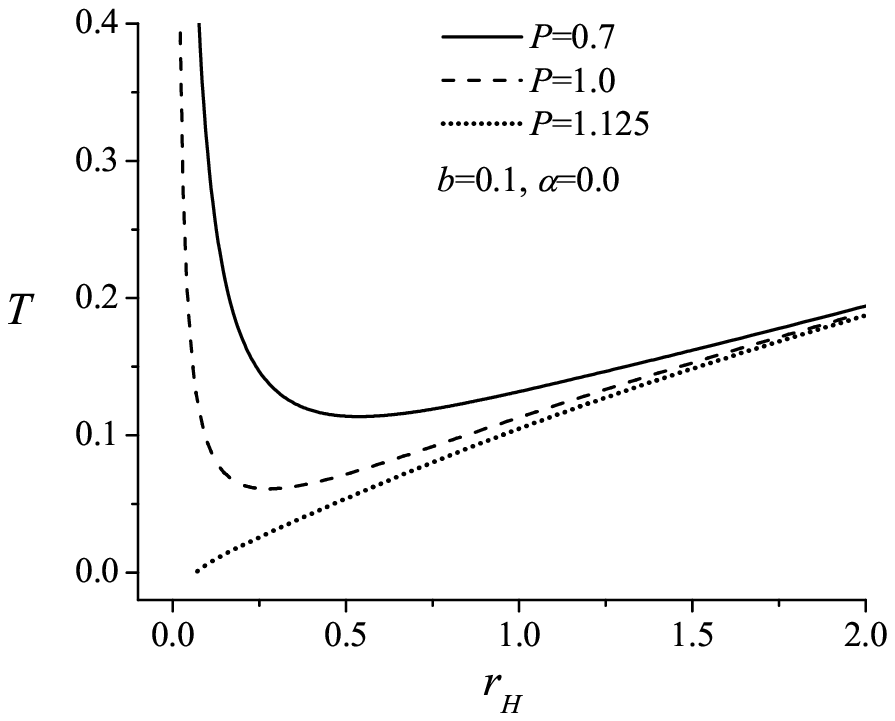}\quad
\includegraphics[width=0.47\textwidth]{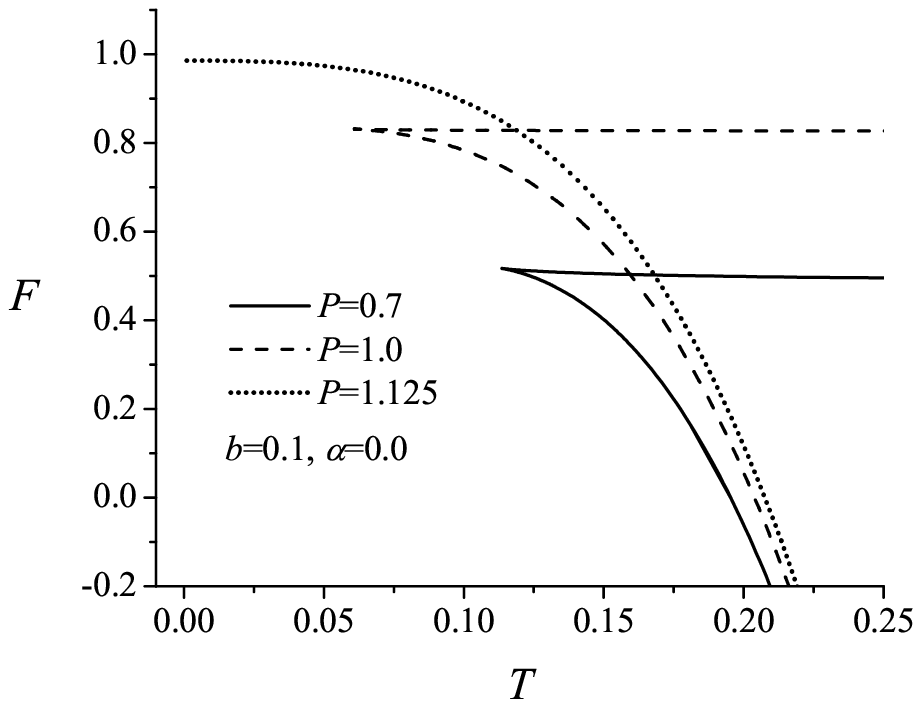}
\caption{The functions $T(r_H)$ and $F(T)$ for sequences of black hole solutions in pure Einstein gravity ($\alpha=0.0$) for a small
value of the Born-Infeld parameter $b=0.1$.}\label{TF_bmalko_alpha0}
\end{figure}

The functions $T(r_H)$ and $F(T)$ for $b=0.1$ are shown in Fig.~\ref{TF_bmalko_alpha0} in the case of pure
Einstein theory which corresponds to $\alpha=0.0$. This case is qualitatively the same as the case when the scalar
field is nontrivial except for large charges which fulfill the inequality $bP^2>1/8$. For these values of $b$ and $P$, when $\alpha=0.0$, the branch of small
black holes disappear because of the occurrence of an extremal black hole (see for example the curve
corresponding to $P=1.125$ in Fig.~\ref{TF_bmalko_alpha0}).

\section{Conclusion}
In the current paper we have constructed numerically new scalar-tensor black hole solutions in AdS space-time coupled to Born-Infeld
nonlinear electrodynamics. The properties of the solutions have been thoroughly studied through a combination of analytical and numerical techniques.
The black hole solutions in the considered class of scalar-tensor theories are determined uniquely by
the asymptotic charges of the black holes -- in our case the mass and the magnetic charge.
An important property is that in comparison to the corresponding solutions in General Relativity,
namely, the Einstein-Born-Infeld-AdS black holes, the black holes considered here have a simpler causal structure -- they have neither inner
nor degenerate horizons.

The thermodynamics and the phase structure in the canonical ensemble of the obtained black holes have also been studied.
Their phase structure is similar to that of the Einstein-Born-Infeld-AdS black holes in GR and the difference is manifested at large charges. This means
that in the case under consideration there are no extremal black hole solutions which actually leads to one additional branch of solutions in the thermodynamical
phase-space structure for large values of the  charge.

The thermodynamical phase-space structure depends on the Born-Infeld parameter $b$ and the magnetic charge $P$. A critical
value $b_{\rm crit}\approx0.6$ has been found numerically. For $b<b_{\rm crit}$ the phase diagram is the same for all values of the magnetic
charge $P$. There are two branches corresponding to unstable, small black holes and locally stable, large black holes. In this case no phase
transitions are possible.

In the case of $b>b_{\rm crit}$, the structure of the phase diagram changes with the variation of the magnetic charge $P$ and three charge intervals are
observed -- small, intermediate, and large charges.
For small and large values of the magnetic charge no phase transitions are possible and the phase diagram has only two branches which
corresponds to two phases: unstable, small black holes and locally stable, large black holes.
The intermediate charges are the most interesting because phase transitions can exist in this case.
The phases are four:
VSBH, SBH, LBH, and VLBH. Among them two are locally stable -- the SBH and the VLBH -- and phase transitions between the two
black holes are possible in some cases. In the intermediate charge subinterval $P^{(2)}_{\rm ph}<P<P^{(2)}_{\rm crit}$ first order phase transition
between SBH and VLBH exists -- for small values of the temperature ($T_{0}<T<T_{\rm ph}$) the SBH are thermodynamically favorable because they have lower
free energy. When we increase the temperature we reach a certain value of the temperature $T_{\rm ph}$ where phase transition to VLBH
occurs and for all temperatures above that value
$T>T_{\rm ph}$ the VLBH are thermodynamically favorable. This phase transition is of first order because the first derivative
of the free energy is discontinuous at that point. An important observation is that for a small intermediate charge subinterval
$P^{(1)}_{\rm ph}<P<P^{(2)}_{\rm ph}$ two phase transitions exist. One of them is the first order phase transition we just described
and the other one is the zeroth order phase transition. In this case for small values of the temperature ($T_{0}<T<T_{\rm min}$) the only locally stable
branch is the VLBH. At temperature $T_{\rm min}$ another locally stable branch appears -- SBH, which have the smallest free energy and dominate the
thermodynamical ensemble. This means that for $T=T_{\rm min}$ phase transition between
VLBH and SBH exists and it is of zeroth order since the free energy is discontinuous at that point. When we increase the temperature the point
$T=T_{\rm ph}$ is reached where the first order phase transition between SBH and VLBH, that was described above, occurs. For $T>T_{\rm ph}$ again
the VLBH dominate the thermodynamical ensemble.
For all the other intermediate charges phase transitions are not possible since the free energy of the VLBH is always the smallest. Similar zeroth and
first order phase transitions for certain charge intervals are present also in the corresponding solutions in General Relativity.

Let us finish with prospects for future work. As in the asymptotically flat case \cite{SYT3}, in AdS spacetimes the scalar-tensor theories with
$\alpha(\varphi)=\beta \varphi$ ($\beta<0$) coupled to nonlinear electrodynamics  admit non-unique black hole solutions \cite{DYKST}.
From AdS/CFT perspectives these non-unique
black hole solutions could be considered as second order phase transitions in the dual field theory living on the boundary.
Detailed considerations will be given in a forthcoming paper \cite{DYKST}. Phase transitions due to non-unique AdS black hole solutions in Einstein-Maxwell-dilaton
gravity with appropriate dilaton coupling function  have been  obtained  quite recently in \cite{Cadoni}. Let us note however that
the situation in the case of scalar-tensor AdS black holes coupled to the nonlinear electrodynamics  is rather different in comparison with the Einstein-Maxwell-dilaton gravity.

\section*{Acknowledgements}
S.Y. would like to  thank
D. Uzunov for the discussions on phase transitions. D.D. would like to thank the DAAD for a scholarship and the Institute for Astronomy and Astrophysics
T\"{u}bingen for its kind hospitality. D.D. would also like to thank J. Kunz, B. Kleihaus, and D. Georgieva for the helpful discussions.

This work was partially supported by the Bulgarian National Science Fund under Grants DO 02-257, VUF-201/06, and by Sofia University Research Fund
under Grant No 074/2009.

\vskip 1cm
\appendix

\section{Black hole thermodynamics in scalar-tensor \\theories}\label{STTERMODYNAMICS}

In this appendix we would like to make some comments on the thermodynamics of the black holes in scalar-tensor theories. The definitions of entropy, temperature, and  mass (energy) of the scalar-tensor black hole are naturally related to the Einstein frame. The reasons are the following.
The Einstein-frame area of the horizon can be interpreted as entropy of the horizon since it is non-decreasing in classical physical processes.
The entropy of the event horizon in the Einstein frame $S_{E}$, for example, is proportional to one fourth of the area of the horizon.  In the Jordan frame, however, the area of the event horizon may decrease in classical processes \cite{Kang} and a generalized definition of entropy $S_{J}$ needs to be introduced. The proper definition is \cite{Kang, MVisser, JKangM, FordRoman}
\begin{equation}
S_{J}=\frac{1}{4G_{*}}\int d^2x \sqrt{-^{(2)}{\tilde g}}F(\Phi).
\end{equation}
This definition is also consistent with the Euclidean and the Noether charge method \cite{JKangM}.

Using relation (\ref{CONFTRANS}) we find that
\begin{equation}
S_{J}=\frac{1}{4G_{*}}\int d^2x \sqrt{-^{(2)} g}=S_{E}=S,
\end{equation}
the  definitions of the entropy in both conformal frames are in agreement.
In the last two equations quantities $^{(2)}{\tilde g}$ and $^{(2)} g$ are the determinants of the induced metrics on the horizon
in the Jordan and in the Einstein frames, respectively.

Further, it can be proved that the standard definitions of black-hole temperature are invariant with respect to conformal transformations that are regular on the event horizon \cite{Jacobson}. Since the conformal transformations (\ref{CONFTRANS}) in the present paper are regular on the horizon
(and everywhere  outside the horizon),  the temperature  in the Einstein frame  coincides with the temperature in the Jordan frame.

The measure of the internal energy of compact objects in STT needs also to be specified. In the general case the
masses
of the black holes in the Einstein frame and in the Jordan frame do not coincide. The mass in the Einstein frame is positive definite,
can only decrease by emission of gravitational (and scalar waves), and possesses other natural energy-like properties unlike the mass in the
Jordan frame. In other words the  mass in the Einstein frame is the true energy of the compact objects within the framework of the STT. For more details on the mass of compact objects in the STT we refer the reader to \cite{Lee,Shapiro,Whinnett,Yazadjiev, SYT1, SYT2}.
For the particular solutions studied in the current paper, however, the scalar field decreases fast enough and the black-hole masses in both conformal frames are the same.

\section{Derivation of the free energy}\label{TD_action}

The  action with surface terms and counterterms   taken into account is given by
\begin{eqnarray}
S= \frac{1}{16\pi } \int_{{\cal M}} d^4x\sqrt{-g}\left(R - 2g^{\mu\nu}\partial_{\mu}\varphi\partial_{\nu}\varphi+ \frac{6}{\ell^2}
+ 4{\cal A}^4(\varphi)L(X)\right)  \nonumber \\ + \frac{1}{8\pi } \int_{\partial {\cal M}}
d^3x \sqrt{-\gamma} \Theta[\gamma] - \frac{1}{8\pi }S_{\rm ct}[\gamma]
\end{eqnarray}
where $\gamma_{\mu\nu}$ is the metric induced on the boundary $\partial {\cal M}$ and $\Theta$ is the trace of the extrinsic curvature
of $\partial {\cal M}$ while $S_{\rm ct}[\gamma]$ is the counterterm action. We have also defined $\ell = \sqrt{-3/\Lambda}$ and
the boundary stress-energy tensor is given by
\begin{eqnarray}
{\cal T}^{\mu\nu}= \frac{2}{\sqrt{-\gamma}} \frac{\delta S}{\delta \gamma_{\mu\nu}}=\frac{1}{8\pi }
\left[\Theta^{\mu\nu}- \Theta \gamma^{\mu\nu} + \frac{2}{\sqrt{-\gamma}} \frac{\delta S_{\rm ct}}{\delta \gamma_{\mu\nu}}\right] .
\end{eqnarray}

If the boundary geometry has an isometry generated by the Killing vector $K$, ${\cal T}_{\mu\nu}K^{\nu}$ is divergence
free which gives the conserved quantity

\begin{eqnarray}
{\cal Q}_{K}=\int_{\sigma} d\sigma^{\mu} {\cal T}_{\mu\nu}K^{\nu}
\end{eqnarray}
associated with the closed surface $\sigma$ in the boundary $\partial {\cal M}$. In the case when $K=\partial/\partial t$, the conserved
quantity is the mass.

For spacetimes with AdS asymptotic the following simple counterterm was proposed in \cite{BKraus} (see also \cite{Mann})

\begin{eqnarray}
S_{\rm ct}=\int_{\partial {\cal M}}d^3x \sqrt{-\gamma} \,\frac{2}{ \ell}\left(1 +\frac{\ell^2}{4}{\cal R}[\gamma]\right)
\end{eqnarray}
where ${\cal R}[\gamma]$  is the Ricci scalar curvature of the boundary metric $\gamma_{\mu\nu}$. With this counterterm
for the asymptotic metric

\begin{eqnarray}\label{AdSasymptotic}
ds^2 \approx -\left(1 - \frac{2M }{r} + \frac{r^2}{\ell^2}\right) dt^2 + \frac{dr^2}{1 - \frac{2M }{r} + \frac{r^2}{\ell^2}}
+ r^2\left(d\theta^2 + \sin^2\theta d\phi^2\right)
\end{eqnarray}
we have up to the leading  order

\begin{eqnarray}
8\pi  {\cal T}_{tt}=\frac{2M}{r\ell} + {\cal O}(r^{-2})
\end{eqnarray}
which gives the mass

\begin{eqnarray}
{\cal Q}_{\xi}=M.
\end{eqnarray}

In other words this shows that the limit $\lim_{r\to \infty} m(r)=M$ of the metric function $m(r)$ (see eq.(\ref{localmass})) has to be identified
with the mass of the solutions with AdS asymptotic.

Our next task is to calculate the Euclidean
action. For definiteness we shall consider the electrically charged case. First to make the action Euclidean,
the time coordinate should be made imaginary by
substitute $t=i\tau$. This makes the metric positively definite, namely

\begin{eqnarray}
ds^2= f(r)e^{-2\delta(r)}d\tau^2 + {dr^2\over f(r) } +
r^2\left(d\theta^2 + \sin^2\theta d\phi^2 \right).
\end{eqnarray}

In order to eliminate the conical singularity at the horizon $r=r_H$, the Euclidean
time coordinate $\tau$ has to be periodic with
period $\beta=1/T$ where $T$ is the Hawking temperature associated with the black hole horizon.  The very Euclidean
action is given by

\begin{eqnarray}
I=I_{\rm bulk} + I_{\Theta} + I_{\rm ct} + I_{\rm cf}
\end{eqnarray}
where

\begin{eqnarray}
I_{\rm bulk}= -\frac{1}{16\pi } \int_{{\cal M}} d^4x\sqrt{g}\left(R - 2g^{\mu\nu}\partial_{\mu}\varphi\partial_{\nu}\varphi+ \frac{6}{\ell^2}
+ 4{\cal A}^4(\varphi)L(X)\right),
\end{eqnarray}

\begin{eqnarray}
I_{\Theta}=   - \frac{1}{8\pi } \int_{\partial {\cal M}}
d^3x \sqrt{\gamma} \Theta[\gamma],
\end{eqnarray}

\begin{eqnarray}
 I_{\rm ct}= \frac{1}{8\pi}\int_{\partial {\cal M}}d^3x \sqrt{\gamma} \,\frac{2}{ \ell}\left(1 +\frac{\ell^2}{4}{\cal R}[\gamma]\right),
\end{eqnarray}

\begin{eqnarray}
I_{\rm cf}= \frac{1}{4\pi } \int_{\partial {\cal M}} d^3x\sqrt{\gamma} n_{\mu}  A_{\nu}F^{\mu\nu}{\cal A}^4(\varphi) \partial_{X}L .
\end{eqnarray}

The bulk integral is on compact region ${\cal M}$ with  boundary $\partial {\cal M}$. The boundary  $\partial {\cal M}$ lays at finite radius
$r=r_{B}$
and has topology $S^{1}\times S^2$ where $S^1$ stands for the periodic Euclidean
time coordinate $\tau$.   The last term $I_{\rm cf}$ is a charge fixing term, in other words we have added this term since we consider a fixed charge
ensemble.

In order to compute the Euclidean
action and to show that $I/\beta=M-TS$ we will need an auxiliary tool, namely the Komar integral

\begin{eqnarray}
{\tilde M}_{B} = - \frac{1}{8\pi } \int_{S^2_{B}}\nabla^{\mu}\xi^{\nu} dS_{\mu\nu}
\end{eqnarray}
where $S^2_{B}$  is a sphere with radius $r_B$ and $\xi$ is the timelike Killing vector. Obviously the Komar integral is divergent for $r_{B}\to
\infty$ but it is finite for finite $r_{B}$.
More precisely, for the AdS asymptotic (\ref{AdSasymptotic}) we have

\begin{eqnarray}\label{KomarMass}
{\tilde M}_{B} = M + \frac{r_{B}^{3}}{\ell^2} + {\cal O}(\frac{1}{r_{B}}).
\end{eqnarray}

Using Stokes theorem we can present the Komar integral as a sum of integral on the horizon ${\cal H}$ and a bulk integral

\begin{eqnarray}
{\tilde M}_{B} = - \frac{1}{8\pi } \int_{{\cal H}}\nabla^{\mu}\xi^{\nu} dS_{\mu\nu}
- \frac{1}{4\pi }\int_{\Sigma} R_{\mu\nu}\xi^{\nu} d\Sigma^{\mu}
\end{eqnarray}
where $\Sigma$ is the 3-dimensional space bounded by $S^2_{B}$ and  ${\cal H}$.

It is well known that the integral on  the horizon gives just

\begin{eqnarray}
 - \frac{1}{8\pi } \int_{{\cal H}}\nabla^{\mu}\xi^{\nu} dS_{\mu\nu}=2TS
\end{eqnarray}
 where $T$ is the Hawking temperature and $S$ is the black hole entropy
which is one fourth of the horizon area. On the other hand using the field equations we find

\begin{eqnarray}
R_{\mu\nu}\xi^{\nu}= \Lambda \xi_{\mu} - 2\partial_{X}L F_{\mu\alpha}F_{\nu}^{\,\,\alpha}\xi^{\nu} +
2{\cal A}^{4}(\varphi)\left[2X\partial_{X}L - L \right]\xi_{\mu}.
\end{eqnarray}
Now we introduce the electric potential $\Phi$ by the equation $\partial_{\alpha}\Phi=-\xi^{\nu}F_{\nu\alpha}$ and taking into account the field
equations for the electromagnetic field, the second term on the right-hand side can be presented in the form

\begin{eqnarray}
  - 2\partial_{X}L F_{\mu\alpha}F_{\nu}^{\,\,\alpha}\xi^{\nu} = 2\partial_{X}L F_{\mu}^{\,\alpha}\partial_{\alpha}\Phi=
  2\nabla_{\alpha}\left(\Phi\partial_{X}L F_{\mu}^{\,\alpha}\right).
\end{eqnarray}
In this way we obtain

\begin{eqnarray}
{\tilde M}_{B} = 2TS -\frac{ \Lambda}{4\pi } \int_{\Sigma} \xi_{\mu}d\Sigma^{\mu}
- \frac{1}{2\pi}\int_{\Sigma} \nabla_{\alpha}\left(\Phi\partial_{X}L F^{\mu\alpha}\right)d\Sigma_{\mu} \nonumber \\
+ \frac{1}{2\pi}\int_{\Sigma}{\cal A}^{4}(\varphi)\left[2X\partial_{X}L - L \right]\xi_{\mu} d\Sigma^{\mu}.
\end{eqnarray}

With the help of the Stokes theorem we find

\begin{eqnarray}
&&- \frac{1}{2\pi}\int_{\Sigma} \nabla_{\alpha}\left(\Phi\partial_{X}L F^{\mu\alpha}\right)d\Sigma_{\mu} =
-\frac{1}{4\pi}\int_{S^2_{B}} \Phi\partial_{X}L F^{\mu\alpha} dS_{\mu\alpha} +
\frac{1}{4\pi}\int_{{\cal H}} \Phi\partial_{X}L F^{\mu\alpha} dS_{\mu\alpha}= \nonumber \\
&&{\cal O}(\frac{1}{r_{B}}) + \frac{\Phi_{H}}{4\pi}\int_{{\cal H}} \partial_{X}L F^{\mu\alpha} dS_{\mu\alpha}=
{\cal O}(\frac{1}{r_{B}}) + 2\Phi_{{\cal H}}Q
\end{eqnarray}
where we have taken into account that at large $r_{B}$ we have $\Phi(r_B)\sim 1/r_B$ and that the electric potential is constant on the horizon.
Also, the electric charge of the black hole is given by

\begin{eqnarray}
Q= \frac{1}{8\pi}\int_{{\cal H}} \partial_{X}L F^{\mu\alpha} dS_{\mu\alpha}.
\end{eqnarray}

Summarizing, we obtained the following relation\footnote{This relation can be considered as some kind of generalized Smarr-like relation.
As it was discussed in \cite{Ras}, in the case of nonlinear electrodynamics (at least simple) Smarr relations do not exist.   }

\begin{eqnarray}\label{Gen_Smarr}
{\tilde M}_{B} = 2TS + 2\Phi_{{\cal H}}Q -\frac{ \Lambda}{4\pi } \int_{\Sigma} \xi_{\mu}d\Sigma^{\mu}
+ \frac{1}{2\pi}\int_{\Sigma}{\cal A}^{4}(\varphi)\left[2X\partial_{X}L - L \right]\xi_{\mu} d\Sigma^{\mu}
\end{eqnarray}
which holds up to terms of order ${\cal O}(r^{-1}_{B})$. Having this important relation we can go back to computation of the Euclidean
action. Using the field equations the bulk term can be presented in the form

\begin{eqnarray}
I_{\rm bulk}= -\frac{\Lambda}{8\pi}\int_{{\cal M}}\sqrt{g}d^4x  - \frac{1}{4\pi}\int_{{\cal M}}{\cal A}^{4}(\varphi) (2X\partial_{X}L -L)\sqrt{g}d^4x.
\end{eqnarray}
Using symmetry along $\xi$ the integration on the periodic coordinate is performed directly and we find

\begin{eqnarray}
I_{\rm bulk}= -\frac{\beta\Lambda}{8\pi}\int_{{\cal M}}\xi_{\mu}d\Sigma^{\mu}
- \frac{\beta}{4\pi}\int_{{\cal M}}{\cal A}^{4}(\varphi) (2X\partial_{X}L -L)\xi_{\mu}d\Sigma^{\mu}.
\end{eqnarray}
At this stage we should make use of the relation (\ref{Gen_Smarr}) which gives

\begin{eqnarray}
\frac{I_{\rm bulk}}{\beta}= \frac{1}{2}{\tilde M}_{B} -TS -\Phi_{{\cal H}}Q
\end{eqnarray}
up to terms of order ${\cal O}(r^{-1}_{B})$. Direct computations of the other Euclidean
terms give the following results

\begin{eqnarray}
&&\frac{I_{\Theta}}{\beta}=\frac{3}{2} M - r_{B} -\frac{3}{2} \frac{r^3_{B}}{\ell^2},\\
&&\frac{I_{\rm ct}}{\beta}= -M + r_{B} + \frac{r^3_{B}}{\ell^2}, \\
&&\frac{I_{\rm cf}}{\beta}= \Phi_{{\cal H}}Q,
\end{eqnarray}
up to terms of order of ${\cal O}(r^{-1}_{B})$. So for the Euclidean
action, up to terms of order of ${\cal O}(r^{-1}_{B})$, we find

\begin{eqnarray}
\frac{I}{\beta}= \frac{1}{2}{\tilde M}_{B} + \frac{1}{2} M - \frac{1}{2} \frac{r^3_{B}}{\ell^2}-TS .
\end{eqnarray}
Finally, making use of (\ref{KomarMass}) we obtain

\begin{eqnarray}
\frac{I}{\beta}= M - TS  + {\cal O}(\frac{1}{r_{B}})
\end{eqnarray}
which shows that in the limit $r_{B}\to \infty$  we indeed have $I/\beta=M-TS$. Let us note that the above considerations
hold for arbitrary nonlinear electrodynamics not only for the Born-Infeld electrodynamics.

\section{Off-shell considerations}\label{TD_offshell}

\begin{figure}[htbp]%
\vbox{ \hfil{\includegraphics[width=0.65\textwidth]{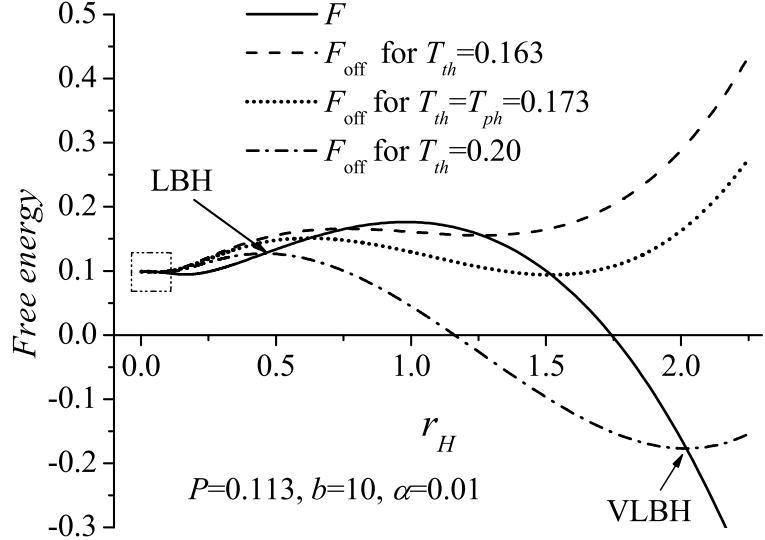} }\hfil}
\caption{%
The free energy $F$ for sequence of black hole solution and the off-shell free energy $F_{\rm off}$ for several different temperatures of the thermostat $T_{th}$
as functions of the radius of the event horizon $r_H$.}\label{Offshell_fig} %
\end{figure}

\begin{figure}[htbp]%%
\vbox{ \hfil{\includegraphics[width=0.70\textwidth]{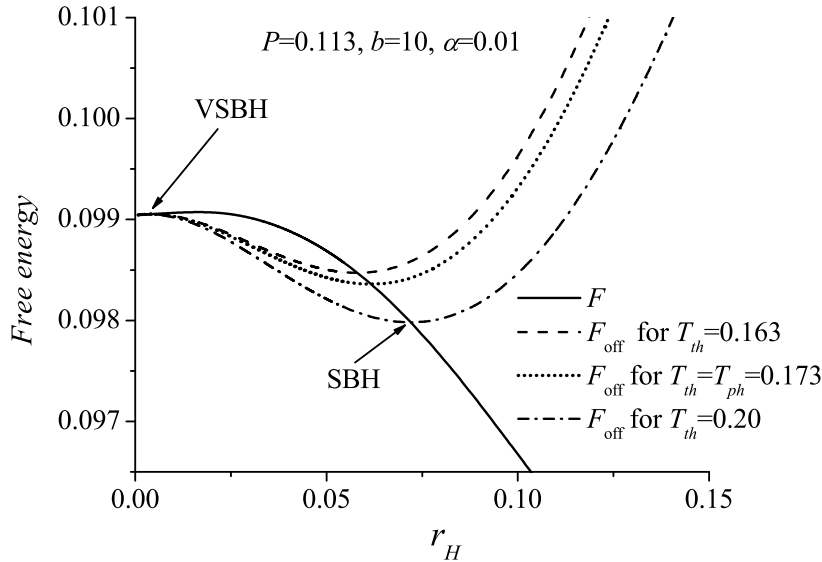} }\hfil}
\caption{%
Magnification of the enclosed region in Fig.~\ref{Offshell_fig}}\label{Offshell_magn}
\end{figure}

Let us consider different black holes, with different masses and entropies\footnote{In the field black-hole thermodynamics it is usually accepted
 to
work with the radius of the event horizon instead of the entropy.}, placed in a thermostat where the role of the thermostat is played by
the Hawking radiation and the appropriate boundary conditions. Then we could ask which black holes would be in equilibrium with the thermostat. The
equilibrium black holes, or in other words the equilibrium states are those for which the off-shell\footnote{This energy is actually the
off-equilibrium energy. The term off-shell is used to emphasize that it is not the natural free energy connected with the solutions. The on-shell
energy, the one related with the solutions, is defined in the same way but $T_{\rm th}$ is substituted with the temperature of the event horizon $T$,
which is different for black holes with different $r_H$.} free energy
\begin{equation}
F_{\rm off}= M-T_{\rm th}S,
\end{equation}
has an extremum. In the above expression $T_{\rm th}$ is the temperature of the thermostat, it is kept fixed, and $S$ and $M$ are, respectively, the
entropy and the mass of a black hole with radius of the event horizon $r_H$. The off-shell free energy for several values of $T_{\rm th}$ is given in
Figs. \ref{Offshell_fig} and \ref{Offshell_magn}.

The equilibrium black holes have temperature of the horizon $T$ that is equal to the temperature of the thermostat $T_{\rm th}$. The different extrema are, actually,
the different phases of the system. The free energy of the different phases, namely, the equilibrium black holes, is plotted with a thick line in Figs. \ref{Offshell_fig}
and \ref{Offshell_magn} and it intersects $F_{\rm off}$ in its extrema. The stable phases are in the minima and the unstable in the maxima of
$F_{\rm off}$. The system is in the phase that realizes a global minimum of $F_{\rm off}$. A phase transition occurs when with the varying of the
temperature of the thermostat $T_{\rm th}$ the system passes from one minimum to another.
The different branches of the equilibrium free energy $F$ as a function of the temperature of the event horizon $T$ (see, for example, the right panel
of  Fig.~\ref{FT_interm_2_2}) can be obtained in the following way. As we know, the thermodynamic parameters,
in our case the temperature and the radius of the black hole, are independent variables. They are related only at equilibrium, i.e. at an extremum
of the off-shell free energy. The condition for an extremum of $F_{\rm off}$ is
\begin{equation}
\left.\frac{\partial F_{\rm off}(r_H)}{\partial r_H}\right|_{ T_{\rm th}=\rm const} = 0
\end{equation}
This relation defines $r_H$ as an implicit function of $T_{\rm th}$ which, in the general case, may have several branches $r_{H,\,a}(T_{\rm th}),\,\,\,\, a=
0,1,2....$. When we take into account that at equilibrium $T=T_{\rm th}$ we obtain the $r_H-T$ diagram (see, for example, the left panel of  Fig.~\ref{FT_interm_2_2}  ). As we can see, the number of the extrema of $F_{\rm off}$ for a given fixed value of $T$ coincides with the number of
branches of the $r_H-T$ diagram for $T=T_{\rm th}$ and the positions of the extrema coincide with the radii of the event horizon of the different branches.

The different branches of the on-shell or the equilibrium free energy (see, for example, the right panel of  Fig.~\ref{FT_interm_2_2}) can be obtained from the off-shell energy in the following
way:
\begin{equation}
F_a(T)=\left.F_{\rm off}(r_{H,\,a}(T_{\rm th}), \,T_{\rm th})\right|_{ T_{\rm th}=T}.
\end{equation}

Let us consider, for example, black holes with $P=0.113$. As can be seen in Fig.~\ref{Offshell_fig} for $T_{\rm th}=0.173$ the minimum corresponding
to SBH and the one corresponding to VLBH are equal so this is the temperature of the phase transition $T_{\rm ph}$.

\begin{figure}[htbp]%
\includegraphics[width=0.48\textwidth]{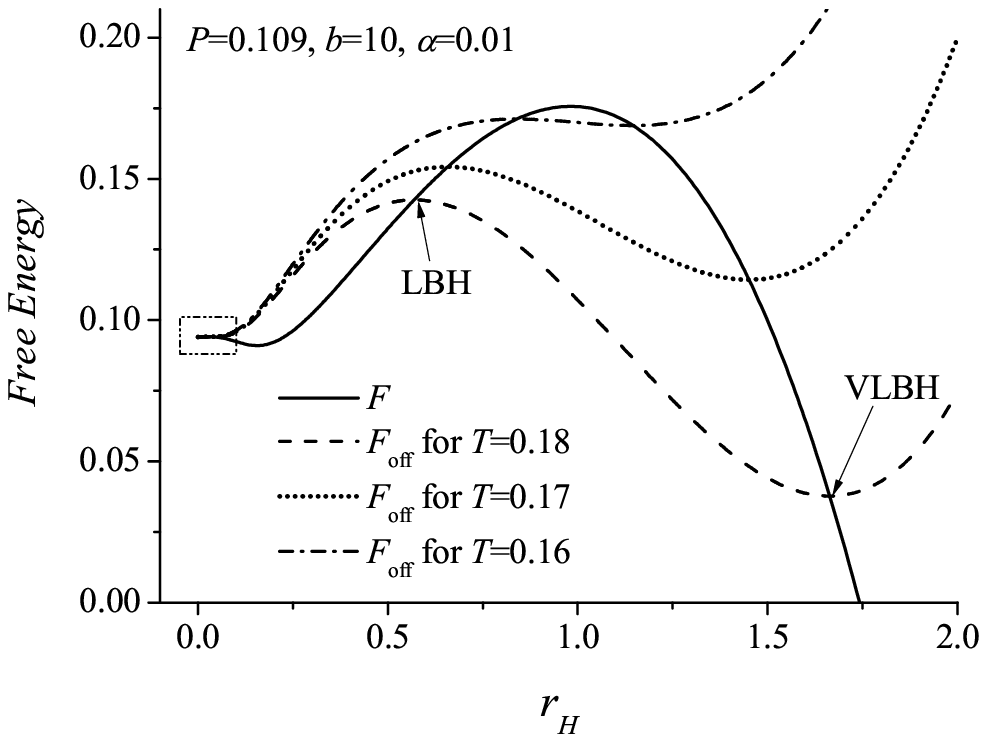}\quad
\includegraphics[width=0.48\textwidth]{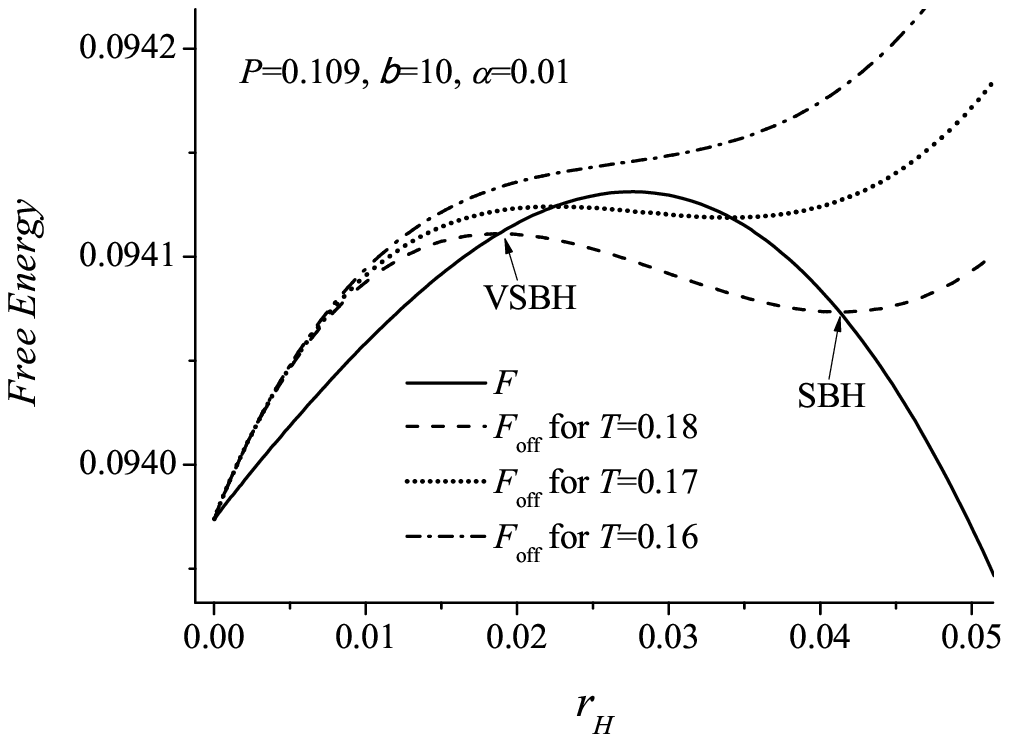}
\caption{The off-shell free energy for the case of the zeroth-order phase transition. The right panel is a magnification of the enclosed
region of the left panel of the figure. }\label{Offshell_zeroth}
\end{figure}

The off-shell formalism can give also a very transparent description of the zeroth-order phase transition that occurs in the subinterval of the
magnetic charge $P^{(1)}_{\rm ph}<P<P^{(2)}_{\rm ph}$. The off-shell free energy is plotted in Fig. \ref{Offshell_zeroth}. The right panel is a magnification
of the enclosed region in the left panel of the same figure. The three curves correspond to temperatures belonging, respectively, to the three intervals
$T_0<T<T_{\rm min}$, $T_{\rm min}<T<T_{\rm ph}$ and $T_{\rm ph}<T$. It can be seen that for $T_0<T<T_{\rm min}$ the off-shell free energy has only two
extrema -- one minimum and one maximum \footnote{Notice that the on-shell free energy intersects the curve of the off-shell free energy only at two points.}. Hence, in that temperature interval only one locally
stable phase is present -- the VLBH. For temperatures in the interval $T_{\rm min}<T<T_{\rm ph}$ a second minimum of the off-shell free energy
corresponding to SBH occurs. It is lower than the minimum corresponding to the VLBH so the SBH is thermodynamically favored. As we have already
mentioned the phase transition from the VLBH to SBH that occurs at $T=T_{\rm min}$ is of zeroth order. As can be seen in Fig. \ref{Offshell_zeroth},
in the last temperature interval  $T_{\rm ph}<T$ the ensemble is again dominated by the VLBH since its minimum is lower than the minimum
corresponding to the SBH and at the point $T=T_{\rm ph}$ first order phase transition occurs.

%%%%%%%%%%%%%%%%%%%%%%%%%%%%%%%%%%%%%%%%%%%%%%%%%%%%%%%%%%%%%%%%%%%%%%%%%%%%%%%


\begin{thebibliography}{9}
%%%%%%%%%%%%%%%%%%%%%%%%%%%%%%%%%%%%%%%%%%%%%%%%%%%%%%%%%%%%%%%%%%%%%%%%%%%%%%%


\bibitem{Maldacena} J. Maldacena, \emph{Adv. Theor. Math. Phys.} \textbf{2}, 231 (1998); arXiv: hep-th/9711200.
\bibitem{Petersen} J. L. Petersen, \emph{Int. J. Mod. Phys.} \textbf{A 14}, 3597 (1999) ;   arXiv: hep-th/9902131
\bibitem{HP} S.~W.~Hawking and D.~N.~Page,
  \emph{Commun. Math. Phys.} {\bf 87}, 577 (1983).



\bibitem{BI} M.~Born~and~L.~Infeld,~\emph{Proc.~R.~Soc.~London}~{\bf A143},~410~(1934).



\bibitem{nle_str1} E. Fradkin and A. Tseytlin, \emph{Phys. Lett.} \textbf{B 163}, 123 (1985).
\bibitem{nle_str2}E. Bergshoeff, E. Sezgin, C. Pope and P. Townsend, \emph{Phys. Lett.} \textbf{B 188}, 70 (1987).
\bibitem{nle_str3}R. Metsaev, M. Rahmanov, and A. Tseytlin, \emph{Phys. Lett.} \textbf{B 193}, 207 (1987).
\bibitem{L} R.~Leigh,~\emph{Mod.~Phys.~Lett.}~{\bf A 4},~2767~(1989).



% grav aspects of NLED
\bibitem{Wil} D. L. Wiltshire,
 \emph{ Phys. Rev. }{\bf  D 38}, 2445 (1988).
\bibitem{Oliveira}H. P. de Oliveira, \emph{Class. Quantum Grav.} \textbf{11}, 1469 (1994).
\bibitem{GR1} G.~Gibbons~and~D.~Rasheed,~\emph{Nucl.~Phys.}~{\bf B 454},~185~(1995).
\bibitem{Ras} D. Rasheed,
  arXiv: hep-th/9702087.
\bibitem{ABG1} E. Ay\'on-Beato and A. Garc\'ia, {\it Phys. Rev. Lett.} {\bf 80}, 5056 (1998)
\bibitem{Novello} M. Novello, S. Perez Bergliaffa and J. Salim, \emph{Class. Quantum Grav.} \textbf{17}, 3821 (2000).
\bibitem{Bronnikov}K. Bronnikov, \emph{Phys. Rev.} \textbf{D 63}, 044005 (2001).
\bibitem{GH} G. W. Gibbons and C. A. R. Herdeiro, \emph{Class. Quant. Grav.} {\bf 18}, 1677 (2001).
\bibitem{TT2}T. Tamaki and T. Torii,\emph{~Phys.~Rev.} {\bf D 64}, 024027 (2001).
\bibitem{BurH}A. Burinskii and S. R. Hildebrandt, \emph{Phys. Rev.} \textbf{D 65}, 104017 (2002).
\bibitem{GS} M.~G\"urses~and~\"O.~Sarioglu,~\emph{Class.~Quantum~Grav.}~{\bf 20},~351~(2003).
\bibitem{Breton2} N. Breton and R. Garcia-Salcedo, Contributed chapter to book on nonlinear electrodynamics edited by CBPF (Brazil); arXiv: hep-th/0702008.



% BI bhs
\bibitem{Garcia} A. Garc\'{i}a, H. Salazar and J. Pleba\~{n}ski, \emph{Nuovo Cimento Soc. Ital. Fis.} \textbf{B 84}, 65 (1984).
\bibitem{Dem} M.~Demianski,
  \emph{Found. Phys. }{\bf 16}, 187 (1986).
\bibitem{CG} G.~Clement~and~D.~Gal'tsov,~\emph{Phys.~Rev.}~{\bf D 62},~124013~(2000).
\bibitem{Tt} T. Tamaki and T. Torii,
 \emph{ Phys. Rev.}  {\bf D 62}, 061501 (2000).
\bibitem{YI} R.~Yamazaki~and~D.~Ida,~\emph{Phys.~Rev.}~{\bf D 64},~024009~(2001).
\bibitem{Breton1} N. Breton; arXiv: gr-qc/0109022
\bibitem{YFBT} S.~Yazadjiev,~P.~Fiziev,~T.~Boyadjiev~and~M.~Todorov,~\emph{Mod.~Phys.~Lett.}~{\bf A 16},~2143~(2001).
\bibitem{FernandoK}S. Fernando and D. Krug, \emph{Gen. Relat. Grav.} \textbf{35}, 129 (2003).
\bibitem{Breton3}N. Breton,  \emph{Phys. Rev.} {\bf D 67}, 124004 (2003).% {\it Born-Infeld black hole in the isolated horizon framework},
%  [arXiv:hep-th/0301254].
  %%CITATION = PHRVA,D67,124004;%%
\bibitem{TTamaki}  T. Tamaki,   \emph{J. Cosm. Astr. Phys.} \textbf{0405}, 004 (2004).
\bibitem{D} T.~Kumar~Dey,~\emph{Phys.~Lett.}~{\bf B 595},~484~(2004).
\bibitem{CPW}R.~Cai,~D.~Pang~and~A.~Wang,~\emph{Phys. Rev.}~{\bf D70},~124034~(2004).
\bibitem{AFG} M. Aiello, R. Ferraro and G. Giribet,
  \emph{Phys. Rev.} {\bf D 70}, 104014 (2004).
%  [arXiv:gr-qc/0408078].
  %%CITATION = PHRV
\bibitem{Y} S.~Yazadjiev,~\emph{Phys.~Rev.}~{\bf D 72},~044006~(2005).
\bibitem{Iran} A.~Sheykhi,~N.~Riazi~and~M.~H.~Mahzoon,~\emph{Phys.~Rev.~}{\bf D 74}, 044025 (2006).
\bibitem{Gao} Xian Gao, \emph{JHEP} \textbf{0711}, 006 (2007).
\bibitem{Yun} Sangheon Yun, arXiv: 0706.2046.
\bibitem{SYT1}I.~Stefanov, S.~Yazadjiev~and M.~Todorov, \emph{Phys.~Rev. }~{\bf D 75}, 084036 (2007).
\bibitem{SYT2}I.~Stefanov,~S.~Yazadjiev~and~M.~Todorov, \emph{Mod.~Phys.~Lett.} {\bf À 22} (17), 1217 (2007).
\bibitem{Sheykhi_topol} A. Sheykhi, \emph{Phys. Lett.} \textbf{B 662}, 7 (2008).



% TD BI bhs
\bibitem{Fern}
S. Fernando,
%  {\it Thermodynamics of Born-Infeld-anti-de Sitter black holes in the grand canonical ensemble},
 \emph{ Phys. Rev.}  {\bf D 74}, 104032 (2006).
%  [arXiv:hep-th/0608040].
  %%CITATION = PHRVA,D74,104032;%%
\bibitem{Sheykhi_higher} A. Sheykhi, N. Riazi, \emph{Phys. Rev.} \textbf{D 75}, 024021 (2007).
\bibitem{TD_BI_BHs} W. A. Chemissany, Mees de Roo, S. Panda, \emph{Class. Quant. Grav.} \textbf{25}, 225009 (2008).
\bibitem{Sheykhi_TD_tpol} A. Sheykhi, \emph{Int. J. Mod. Phys. } \textbf{D 18}, 25 (2009); arXiv: 0801.4112.

%STT
\bibitem{Will} C. M. Will, \emph{``The Confrontation between General Relativity and Experiment''}, \emph{Living Rev. Relativity} \textbf{9}, 3 (2006), URL (cited on
    10.11.2009): http://www.livingreviews.org/lrr-2006-3, arXiv:gr-qc/0103036.

\bibitem{SYT3}I. Stefanov, S. Yazadjiev and M. Todorov, \emph{Mod. Phys. Lett.} \textbf{A 23} (34), 2915 (2008), arXiv: 0708.4141.



%Phase trans
\bibitem{Myung3} Yun Soo Myung, Yong-Wan Kim and Young-Jai Park,  \emph{Phys. Rev.} \textbf{D 78}, 084002 (2008); arXiv: 0805.0187.

\bibitem{EmparanJohnson} A. Chamblin, R. Emparan, C. V. Johnson, R. C. Myers,   \emph{Phys.Rev.} \textbf{D 60}, 064018 (1999); arXiv:
    hep-th/9902170.
\bibitem{EmparanJohnson1}A. Chamblin, R. Emparan, C. V. Johnson, R. C. Myers, \emph{Phys.Rev.} \textbf{D 60},  104026  (1999); arXiv:hep-th/9904197.
\bibitem{Myung1} Yun Soo Myung, \emph{Mod. Phys. Lett.} \textbf{A 23}, 667 (2008).
\bibitem{Myung2} Yun Soo Myung, \emph{Phys. Lett.} \textbf{B 663}, 111 (2008).

\bibitem{Dey1} T. Kumar Dey, S. Mukherji, S. Mukhopadhyay and S.Sarkar,\emph{ JHEP} \textbf{0704},014 (2007).
\bibitem{Dey2} T. Kumar Dey, S. Mukherji, S. Mukhopadhyay and S.Sarkar,\emph{ JHEP} \textbf{0709},026 (2007).
\bibitem{Cho}Cho Y. M. Cho and Ishwaree P. Neupane,    \emph{Phys. Rev.} \textbf{D 66 }, 024044 (2002).
\bibitem{MGH} H. A. Gonzalez, M. Hassaine and C. Martinez, \emph{Phys. Rev.}
\textbf{D 80}, 104008 (2009); \emph{CECS-PHY-09/08 }; arXiv:0909.1365 [hep-th].
\bibitem{MannStotyn} S. Stotyn and R. Mann, \emph{Phys. Lett.} \textbf{B 681},  472  (2009); arXiv:0909.0919 [hep-th].



\bibitem{GR1} G.~Gibbons~and~D.~Rasheed,~\emph{Nucl.~Phys.}~{\bf B454},~185~(1995).
\bibitem{GR2} G.~Gibbons~and~D.~Rasheed, \emph{Phys.Lett.} \textbf{B365}, 46 (1996).
\bibitem{Salazar} Humberto Salazar I., Alberto Garc\'{\i}a D. and Jerzy Pleba\'{n}ski, \emph{J.~Math. Phys.} {\bf 28}(9), 2171 (1987).
\bibitem{KB} K. Bronnikov, \emph{Phys.~Rev}. {\bf D63}, 044005 (2001).
\bibitem{Ferrara} Paolo Aschieri, Sergio Ferrara and Bruno Zumino, \emph{Riv.Nuovo Cim.} \textbf{31}, 625 (2008); arXiv: 0807.4039 [hep-th] (2008).




\bibitem{gavurin} M.~K. Gavurin, %Nonlinear functional equations and
                 %continuous analogues of iterative methods,
                 Izvestia VUZ, \emph{Matematika} {\bf 14}(6), 18--31 (1958) (in Russian) (see also
                 \emph{Math. Rev.} {\bf 25}, 1380  (1963)).
\bibitem{jidkov} E.~P. Jidkov, G.~I. Makarenko and I.~V. Puzynin,
                 %Continuous analog of Newton's method in nonlinear problems of
                 %Physics,
                 in {\it Physics of Elementary Particles and Atomic
                 Nuclei} (JINR, Dubna, 1973) vol. 4, part I, 127--166 (in
                 Russian), English translation: {\it American Institute of
                 Physics}, p.53.

\bibitem{JGGA2} T. L. Boyadjiev, M. D. Todorov, P. P. Fiziev and S. S. Yazadjiev,  \emph{J. Comp. Appl. Math.} \textbf{145(1)}, 113 (2002);    arXiv:
    math/0004108 [math.NA].


\bibitem {BKraus}V. Balasubramanian and P. Kraus, \emph{Commun. Math. Phys.} \textbf{208}, 413 (1999).

\bibitem{Arcioni}G.~Arcioni and E.~Lozano-Tellechea, \emph{Phys.Rev.} {\bf D 72}, 104021 (2005).


\bibitem{Vega}H. J. de Vega and N. S'anchez, \emph{Nucl. Phys. }\textbf{B 625},  409 (2002), arXiv: astro-ph/0101568.
\bibitem{Jagla} S. Bustingorry, E. A. Jagla and J. Lorenzana, \emph {Acta Materialia}, {\bf 53}, Issue 19,  5183 (2005);
arXiv: cond-mat/0509489 (see also arXiv: cond-mat/0307134) .

\bibitem{Maslov} V. Maslov, \emph{Theor.  and Math. Phys.} \textbf{150}, 102 (2007).
\bibitem{DYKST} D. Doneva, S. Yazadjiev, K. Kokkotas, I. Stefanov and M. Todorov, paper in preparation
\bibitem{Cadoni} M. Cadoni, G. D'Appollonio and P. Pani, \emph{J. High Energy
Phys.} \textbf{03} (2010) 100; arXiv: 0912.3520 [hep-th].

%%%%%%%%%%%%%%%%%%%%%%
\bibitem{Kang} G. Kang,~Phys.~Rev.~{\bf D54},~7483~(1996).
\bibitem{MVisser} M.~Visser,~Phys.~Rev.~{\bf D48},~583~(1993).
\bibitem{JKangM} T. Jacobson, G. Kang and R. Myers, ~Phys.~Rev.~{\bf D52},~3518~(1995).
\bibitem{FordRoman} L.~H.~Ford~and~Thomas~A.~Roman,~Phys.~Rev.~{\bf D64},~024023~(2001).


\bibitem{Jacobson} T.~Jacobson~and~G.~Kang,~Class.~Quant.~Grav. {\bf 10} L201 (1993).

\bibitem{Lee}D.~L.~Lee,~Phys.~Rev.~{\bf D10},~2374~(1974).
\bibitem{Shapiro}~M.~A.~Scheel,~S.~L.~Shapiro,~and~S.~A.~Teukolsky,~Phys.~Rev.~{\bf D~51}, 4208 (1995).
\bibitem{Whinnett} A.~W.~Whinnett,~Class.~Quantum~Grav.~{\bf 16}, 2797 (1999).
\bibitem{Yazadjiev} S.~S.~Yazadjiev,~Class.~Quantum~Grav.~{\bf 16},  L63  (1999).

\bibitem{Mann} R.B. Mann, \emph{Phys. Rev.} \textbf{D 60}, 104047 (1999);  \emph{Phys. Rev. }\textbf{D 61}, 084013 (2000).


\end{thebibliography}
\end{document}